\documentclass[aps,prd,superscriptaddress,onecolumn,showpacs,nofootinbib,preprintnumbers,noshowpacs]{revtex4}

\usepackage{amsmath,amssymb,hyperref,subfigure,xspace}
\usepackage{mciteplus,color,mathtools,graphicx}
\usepackage[english]{babel}
\usepackage{todonotes}
\usepackage{dsfont}
\usepackage[percent]{overpic}
\usepackage{amsmath}
\usepackage{rotating}

\definecolor{colZer}{rgb}{0.76,0.43,0.81}
\definecolor{colOne}{rgb}{0.0,0.6,0.97}
\definecolor{colTwo}{rgb}{0.88,0.43,0.27}
\definecolor{colThr}{rgb}{0.23,0.63,0.29}

\newcommand{\wigner}{\zeta}

\newcommand{\braket}[2]{\left\langle#1|#2\right\rangle}
\newcommand{\HellH}[1]{\ensuremath{H^{#1}}}
\newcommand{\hellH}[1]{\ensuremath{h^{#1}}}
\newcommand{\ie}{\textit{i.e.}\ }
\newcommand{\eg}{\textit{e.g.}\ }
\newcommand{\cf}{\textit{cf.}\ }

\newcommand{\Props}[2]{\ensuremath{X_s^{#2}(#1)}}
\newcommand{\Ps}[1]{\Props{#1}{}}

\newcommand{\diff}{\mathrm{d}}

\newcommand\tikznode[3][]%
   {\tikz[remember picture,baseline=(#2.base)]
      \node[minimum size=0pt,inner sep=0pt,#1](#2){#3};%
   }
\tikzset{>=stealth}

\allowdisplaybreaks

\begin{document}

\graphicspath{{figs/}}

\title{Dalitz-plot decomposition for three-body decays}

\newcommand{\cern}{CERN, 1211 Geneva 23, Switzerland}
\newcommand{\jlab}{Theory Center,
Thomas  Jefferson  National  Accelerator  Facility,
Newport  News,  VA  23606,  USA}
\newcommand{\ucm}{Departamento de F\'isica Te\'orica, Universidad Complutense de Madrid, 28040 Madrid, Spain}
\newcommand{\icn}{Instituto de Ciencias Nucleares,
Universidad Nacional Aut\'onoma de M\'exico, Ciudad de M\'exico 04510, Mexico}
\newcommand{\ceem}{Center for  Exploration  of  Energy  and  Matter,
Indiana  University,
Bloomington,  IN  47403,  USA}
\newcommand{\indiana}{Physics  Department,
Indiana  University,
Bloomington,  IN  47405,  USA}
\newcommand{\ect}{European Centre for Theoretical Studies in Nuclear Physics and related Areas (ECT$^*$) and Fondazione Bruno Kessler, Villazzano (Trento), I-38123, Italy}
\newcommand{\genova}{INFN Sezione di Genova, Genova, I-16146, Italy}
\newcommand{\icsup}{Institute of Computer Science, Pedagogical University of Cracow, 30-084 Krak\'ow, Poland}

\newcommand{\edi}{School of Physics and Astronomy, University of Edinburgh, EH9 3FD Edinburgh, United Kingdom}

\author{M.~Mikhasenko}
\email{mikhail.mikhasenko@cern.ch}
\affiliation{\cern}

\author{M.~Albaladejo}
\affiliation{\jlab}

\author{\L.~Bibrzycki}
\affiliation{\indiana}
\affiliation{\jlab}
\affiliation{\icsup}

\author{C.~\surname{Fern\'andez-Ram\'irez}}
\affiliation{\icn}

\author{V.~Mathieu}\affiliation{\ucm}

\author{S.~Mitchell}\affiliation{\edi}

\author{M.~Pappagallo}\affiliation{\edi}

\author{A.~\surname{Pilloni}}
\affiliation{\ect}
\affiliation{\genova}

\author{D.~Winney}
\affiliation{\ceem}
\affiliation{\indiana}

\author{T.~Skwarnicki}
\affiliation{Syracuse University, Syracuse, NY 13244, USA}

\author{A.~P.~Szczepaniak}
\affiliation{\jlab}
\affiliation{\ceem}
\affiliation{\indiana}

\begin{abstract}
We present a general formalism to write the decay amplitude for multibody reactions
with explicit separation of the rotational degrees of freedom,
which are well controlled by the spin of the decay particle,
and dynamic functions on the subchannel invariant masses,
which require modeling.
Using the three-particle kinematics we demonstrate the proposed factorization, named the Dalitz-plot decomposition.
The Wigner rotations, which are subtle factors needed by the isobar modeling in the helicity framework,
are simplified with the proposed decomposition.
Consequently, we are able to provide them in an explicit form suitable for the general case of arbitrary spins.
The only unknown model-dependent factors are the isobar lineshapes that describe the subchannel dynamics.
The advantages of the new decomposition are shown through three examples relevant for the recent discovery
of the exotic charmonium candidate $Z_c(4430)$, the pentaquarks~$P_c$, and the intriguing $\Lambda_c^+\to pK^-\pi^+$ decay.
\end{abstract}

\preprint{JLAB-THY-19-3070}


\maketitle
\section{Introduction}
Partial-wave decomposition of reaction amplitudes is widely used in
the analysis of both fixed target (\eg COMPASS, VES, CLAS, GlueX),
and collider (\eg LHCb, BESIII, Belle, BaBar) experiments. It is the most powerful way to account for spin and parity, $J^P$, of various contributions, thus is required in determinations of the quantum number for newly observed resonances. It also provides for the most sensitive way of distinguishing exotic hadrons, including the $XYZ$ states and pentaquark candidates in the heavy quarkonium sector, from usually large contributions by ordinary mesons and baryons.
To establish the existence of a resonance in a given partial wave, it is desired to have a representation of the reaction amplitude
consistent with the $S$-matrix principles of unitarity, analyticity and Lorentz invariance.
This is nontrivial when dealing with particles with spin, which introduce kinematical singularities and (pseudo)threshold relations between partial waves.
Amplitude analysis in the context of the $S$-matrix constraints has been extensively studied in the past using both
covariant~\cite{Anisovich:2006bc, Filippini:1995yc, Chung:2007nn} and noncovariant methods~\cite{Zemach:1968zz, Jacob:1959at, Chung:1993da, cookbook}.
When several particles with spin are involved, the noncovariant approach is more practical, because spin is universally accounted for through the simple Wigner $D$-functions.
In this paper, we take a step to simplify amplitude construction and discuss a convenient framework which incorporates dynamic subchannel resonances for a multiparticle decay.
We present a universal amplitude formula which describes the decay of an arbitrary spin state to three particles, each also with arbitrary spin.
Specifically, we write the amplitudes in a factorized form to separate the dependence on the angles that characterize the orientation of the final-state particles
(and thus the information about the polarization of the parent particle) from the Dalitz-plot variables that encode the information on the intermediate resonances in the multiparticle final state.
We do not focus on details of the two-particle dynamics merely giving an example of the simplest parametrization, however,
we stress that additional constraints from kinematical singularities outside of the physical region of the decay
(\eg see Refs.~\cite{Mikhasenko:2017rkh, Pilloni:2018kwm}) can be applied in our framework
and would lead to more complicated lineshape functions.
Furthermore, two-body unitarity constraints (\eg see Refs.~\cite{Aitchison:1966lpz, Aitchison:1979fj, Niecknig:2015ija, Mikhasenko:2016mox}) can be used to confine uncertainties of
the two-body dynamics. While the latter lies outside of the scope of this paper, our framework provides a convenient basis for this investigation.

The rest of the paper is organized as follows.
The details of the amplitude construction are discussed in Sec.~\ref{sec:dalitz.plot.decomposition}.
In Sec.~\ref{sec:examples} the formalism is illustrated with three specific examples,
namely, $\Lambda_c^+\to p K^- \pi^+$, $\overline{B}^0\to \psi [\to \mu^+ \mu^-] \pi^+ K^-$, and $\Lambda_b^0\to J/\psi [\to \mu^+ \mu^-] p K^-$.
These reactions are relevant in exotic hadron searches and/or carry particular complications due to spin. All the necessary derivations are summarized in the appendices, where we also compare our method to other approaches.

\section{Dalitz-plot decomposition} \label{sec:dalitz.plot.decomposition}

\begin{figure}
  \centering
  \includegraphics[]{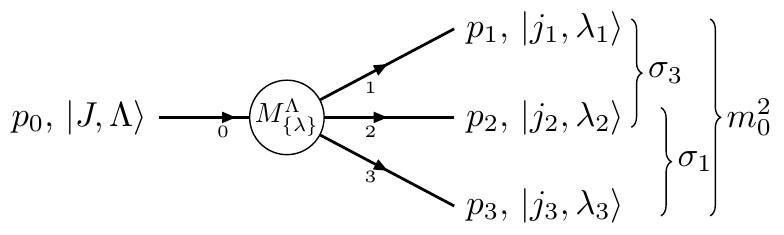}
  \caption{
  Diagram for the three-body decay of a particle with spin $J$ to particles labeled $1,2,3$ with $j_1$, $j_2$, $j_3$ spins, respectively.}
  \label{fig:general.1to3}
\end{figure}

We focus on three-body decays, labeled as $0 \to 1\,2\,3$ as shown in Fig.~\ref{fig:general.1to3}, where particles have arbitrary spin.
The particles $1$, $2$, and $3$ can decay further, however, we assume that their lifetimes are large enough so that the interaction between their decay products and the other particles can be neglected.
In this case, the subsequent decay factors out of the $0 \to 1\,2\,3$ process.
This holds for particles that are stable under the strong interaction ($\pi^0$, $D$, \dots), as well as for narrow resonances such as $J/\psi$, $\phi$, $\eta'$, \dots.
For simplicity we omit isospin indices and comment on the treatment of identical particles later in the text.
The reference coordinate system is fixed in the rest frame of the decaying particle.
The configuration of the momenta in this frame is referred to as a \textit{space-fixed} center-of-momentum configuration (CM).
The three-momenta of the decay products span a plane; therefore, it is convenient to also consider an additional configuration.
A specific event is said to be in an \textit{aligned} configuration if the decay-product plane coincides with the $xz$ plane of the coordinate system.
Any event can be brought into the space-fixed configuration for the aligned one by an overall rotation determined by a set $(\alpha,\beta,\gamma)$ of Euler angles to be specified below~\cite{Ascoli:1974sp}.
The dependence of the reaction differential decay width on these angles is determined by the particle-$0$ spin-density matrix, and, for example disappears in the unpolarized case.
In general, the choice of coordinates of the space-fixed frame is arbitrary, however, the polarization matrix is simplified
(\eg, it is diagonal for a spin-$1/2$ particle) when the $z$ axis points in the direction of the polarization.
For production of particles in colliding beams, the longitudinal polarization is suppressed due to parity conservation of the strong interaction~\cite{Bunce:1976yb}.
Therefore, for polarization studies, it is convenient to choose the $z$ axis parallel to $\vec p_{\mathrm{beam}}\times \vec p_0$~\cite{Hrivnac:1994jx,Aaij:2013oxa}.
The transverse direction is preserved when the system is boosted to the rest frame of particle $0$.
The $xz$ plane is specified by requiring that it contains $\vec p_{\mathrm{beam}}$.
As an alternative to the transverse frame, there are several possible longitudinal frames commonly used for polarization studies of charmonium~\cite{Faccioli:2010kd}.
Throughout the paper we use \textit{active} transformations, \ie the coordinates of the reference frames are fixed while the particle four-vectors change under boosts or rotations.

\begin{figure}
  \includegraphics[width=0.8\textwidth]{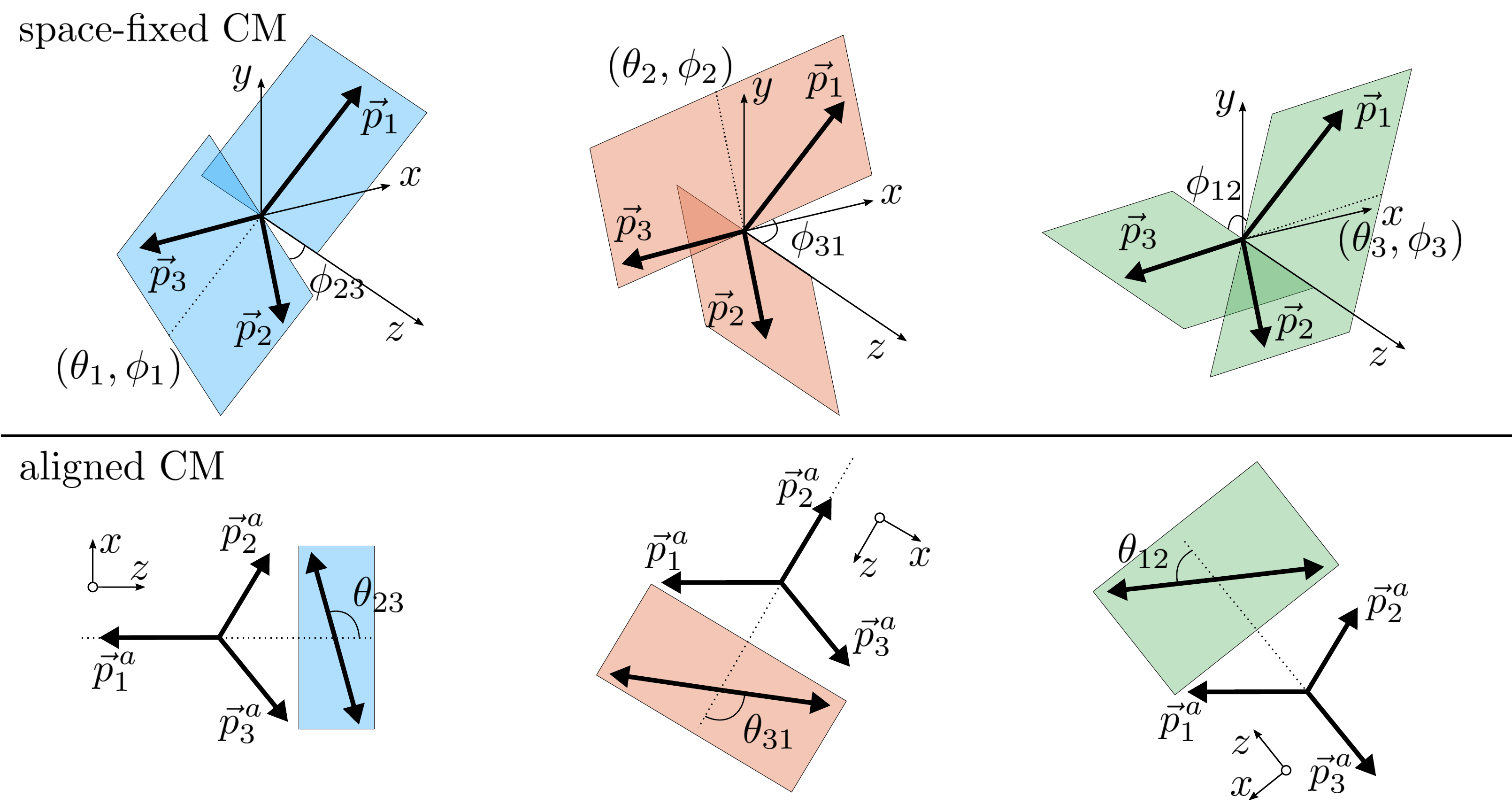}
  \caption{Three different choices of the Euler rotations that lead to different aligned center-of-momentum (aligned CM) frames.
  The upper row shows the measured space-fixed frame. The coordinate axes are fixed by the external conditions, such as the production mechanism and the definition of the polarization matrix.
  For the three cases, $(ijk) \in \{(123),(231),(312)\}$, the angles with a single index $(\theta_k,\phi_k)$ provide the direction of $-\vec p_k$, while the angles with the double index $(\theta_{ij},\phi_{ij})$
  give the direction of $\vec p_i'$ vector in the isobar-$k$ helicity frame (\ie $(ij)$-rest frame).
  The lower row shows the orientation of the vectors in the aligned CM, depicting the momenta of particles $i$, and $j$ in the $(ij)$ rest frame.}
  \label{fig:Euler.angles}
\end{figure}

In the following, we denote the transition amplitude
for an initial state with spin $J$,
and spin projection $\Lambda$ quantized along the $z$ axis in the space-fixed frame by $M^{\Lambda}_{\{\lambda\}}$.
Individual spins and helicities of the three particles in the final state are denoted by $j_i$ and $\lambda_i$, respectively, and collectively by $\{\lambda\} \equiv (\lambda_1,\lambda_2,\lambda_3$).
The amplitude $M^{\Lambda}_{\{\lambda\}}$ can be written,
\begin{equation} \label{eq:proposal}
M^{\Lambda}_{\{\lambda\}} = \sum_{\nu} D^{J*}_{\Lambda,\nu}(\alpha,\beta,\gamma) O^{\nu}_{\{\lambda\}},
\end{equation}
where the Wigner $D$-function stands for the $(2J+1)$-dimensional spinor representation of the rotation group (see \eg Ref.~\cite{Chung:1971ri}),
\begin{equation} \label{eq:wigner.funciton}
D^{J}_{\Lambda,\nu}(\alpha,\beta,\gamma) = \left\langle J,\Lambda | e^{-i\alpha J_z} e^{-i\beta J_y} e^{-i\gamma J_z} | J,\nu \right\rangle.
\end{equation}
This rotation moves the
momenta of the final-state particles from the aligned configuration ($\vec p^{\,a}_1$, $\vec p^{\,a}_2$, $\vec p^{\,a}_3$) to the
measured one, ($\vec p_1$, $\vec p_2$, $\vec p_3$).
In this aligned configuration $-\vec p^{\,a}_1$ is oriented along the $z$ axis and ($\vec p^{\,a}_1$, $\vec p^{\,a}_2$, $\vec p^{\,a}_3$) lie in the $xz$ plane.
The vectors in the measured (space-fixed) configuration are obtained by first rotating the aligned configuration about the $z$ axis by $\gamma$, followed by  rotations by $\beta$ and $\alpha$ about $y$ and $z$, respectively,
where $\beta$ and $\alpha$ are the polar and azimuthal angles of the measured direction of the $-\vec p_1$.
The angle $\gamma$ is the azimuthal angle between the space-fixed $y$ axis and the normal to the particles plane given by
$\vec p_2 \times \vec p_3$,
once $\vec p_1$ has been aligned with the $-z$ axis (see the first column in Fig.~\ref{fig:Euler.angles}, with $\alpha = \phi_1$, $\beta = \theta_{1}$, and
$\gamma = \phi_{23}$).
The index $\nu$ corresponds to the component the spin of the particle-$0$ quantized along the direction opposite to particle $1$.
The Euler angles appear naturally in a sequential decay of the particle-$0$ into an \textit{isobar} (two-particle subsystem) and a spectator (particle $1$), followed by the isobar decay to particles $2$ and $3$.
$\Omega=(\alpha,\beta)$ is the spherical angle determining the direction of the
isobar motion in the space-fixed CM, and $\gamma$ is the azimuthal angle of the relative momentum between $2$ and $3$ in the isobar helicity frame,
obtained from the space-fixed CM with inverse rotation by $\Omega$ and a boost along $z$ axis.
The amplitude $O^{\nu}_{\{\lambda\}} = O^{\nu}_{\{\lambda\}}(\{\sigma\})$ describes the transition to the three-particle final state
in the aligned configuration, for which the relative motion between the particles is completely specified by Lorentz-invariant variables, $\{\sigma\}$.
In the following, we refer to it as the \textit{Dalitz-plot function}.
For $0 \to 1\,2\,3$ decay we employ the Mandelstam variables:
$\sigma_1 = (p_2+p_3)^2$, $\sigma_2 = (p_1+p_3)^2$ and $\sigma_3 = (p_1+p_2)^2$,
related by
\begin{equation*}
  \sigma_1 + \sigma_2 + \sigma_3 = \sum_{i=0}^{3} m_i^2,
\end{equation*}
where $m_i$ are the masses of the particles. In terms of the Dalitz-plot function the differential cross section reads,
\begin{equation} \label{eq:decay}
  \diff \sigma / \diff \Phi_3 = N
  \sum_{\Lambda,\Lambda'} \rho_{\Lambda\Lambda'}
  \sum_{\nu,\nu'} D^{J*}_{\Lambda,\nu}(\alpha,\beta,\gamma) D^{J}_{\Lambda',\nu'}(\alpha,\beta,\gamma)
  \sum_{\{\lambda\}}O^{\nu}_{\{\lambda\}} O^{\nu'*}_{\{\lambda\}},
\end{equation}
where $N$ is an overall normalization factor, and $\rho$ is the spin-density matrix of the decaying particle.
It is clear that in the unpolarized case, when $\rho_{\Lambda\Lambda'} \sim \delta_{\Lambda\Lambda'}$, the dependence on $\alpha$, $\beta$, and $\gamma$ drops out.
Conversely, when one integrates over the Euler angles, the remaining distribution is not sensitive to the polarization.

The amplitude $M^\Lambda_{\{\lambda\}}$ can be written as a sum of three terms, each one defining its own aligned configuration,
\begin{equation} \label{eq:isobar}
M_{\{\lambda\}}^{\Lambda} =
 M_{\{\lambda\}}^{(1), \Lambda}
     + M_{\{\lambda\}}^{(2), \Lambda}
      + M_{\{\lambda\}}^{(3), \Lambda}.
\end{equation}
Each term describes a two-particle partial-wave (isobar) sum
labeled in the superscript with the index of the spectator particle to distinguish the three types of isobars.
The isobar can alternatively be identified by the indices of the two particles it decays into. In the following we use both notations:
the single-index notation is used to specify the isobar angles in the CM frame,
while the double-index notation is used for the angles of isobar-decay products (see examples in Fig.~\ref{fig:Euler.angles}).

\begin{figure}[t]
  \includegraphics[width=\textwidth]{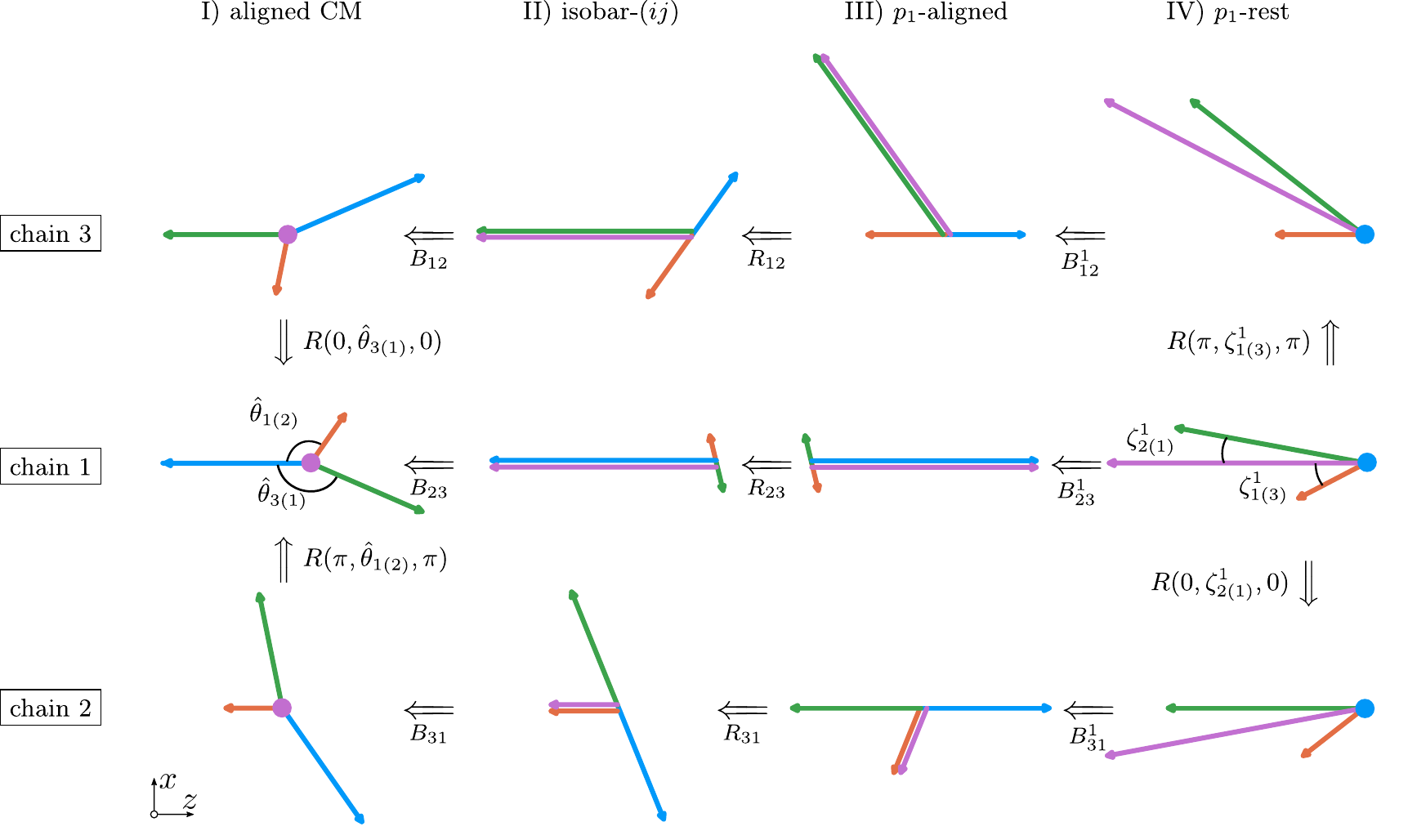}
  \caption{
    Transformations of the aligned configurations of momenta in the decay
    ${\color{colZer}p_0(\text{purple})}\to
      {\color{colOne}p_1(\text{blue})}\,
      {\color{colTwo}p_2(\text{orange})}\,
      {\color{colThr}p_3(\text{green})}$.
    The rows correspond to the decay chains $3(12)$, $1(23)$, $2(31)$, respectively.
    The columns are different frames for each chain $k(ij)$:
      $(\text{I})$ the aligned CM with $\vec p_k$ pointing to $-z$ direction,
      $(\text{II})$ vectors are boosted to the isobar $k$ rest frame, where $\vec p_i + \vec p_j=\vec 0$,
      $(\text{III})$ the same configuration as before, but with $\vec p_1$ aligned with $z$,
      and $(\text{IV})$ vectors are boosted to particle-$1$ rest frame to show how the Wigner angles arise.
    The black arrows indicate the transformations, with self-explanatory indices.
    The clockwise rotations about the $y$ axis are implemented with $R(\pi,\theta,\pi)$,
    and the plane is flipped by $R_z(\pi)$ before and after the $y$-rotation (see Eq.~\eqref{eq:R.clockwise}).
  }
  \label{fig:proton.helicity}
\end{figure}

In practical cases, one or more terms in Eq.~\eqref{eq:isobar} can be neglected if no sizable interaction happens in that subchannel, \eg as in $\pi^+\pi^+$.
Schematically, the individual amplitudes, $M^{(i)}$ are given by
the product of two subsequent two-body decay amplitudes.
The first one,
\begin{equation} \label{eq:h1}
  n_J\,
  D_{\Lambda,\tau-\lambda_k}^{J*}(\Omega_{k})\,
  H^{0\to (ij),k}_{\tau,\lambda_k},
\end{equation}
describes the decay of the particle-$0$ to the isobar $(ij)$ and the spectator $k$.
Here, $n_J=\sqrt{2J+1}$ is a common normalization factor,
$\tau$ and $\lambda_k$ are the helicities of the isobar and the spectator particle, respectively in the space-fixed CM.
The second one,
\begin{equation}\label{eq:h2}
  n_s\,
  D_{\tau,\lambda_i'-\lambda_j'}^{s*}(\Omega_{ij})\,
  H^{(ij)\to i,j}_{\lambda_i',\lambda_j'}\,
\end{equation}
describes the decay of the isobar, with $\lambda_i'$ and $\lambda_j'$ denoting the helicities of the decay products, $n_s=\sqrt{2s+1}$.
We note that the two amplitudes given above are evaluated in different frames: Eq.~\eqref{eq:h1} is evaluated
in the space-fixed CM, while Eq.~\eqref{eq:h2} is computed in the isobar helicity frame.
The boost that relates the two frames affects the helicities of particles $i$ and $j$ as discussed below.
In Eqs.~(\ref{eq:h1},\ref{eq:h2}), $\Omega$ denotes a pair of spherical angles, and the $D$ function reads $D(\Omega) = D(\phi,\theta,0)$.
For each term in Eq.~\eqref{eq:isobar}, these angles are tied to a different aligned configuration.
The angles associated with the isobar in channel $k$ are defined in the space-fixed CM.
$\Omega_k$ is the spherical angle of the momentum of the isobar, \ie $\vec p_i+ \vec p_j$ (see Fig.~\ref{fig:Euler.angles}),
while the spherical angle $\Omega_{ij}$ specifies the direction of motion of particle $i$ in the isobar helicity frame.
The latter is obtained from the space-fixed CM by applying a rotation inverse to $R(\Omega_k)$ and a boost along the $z$ axis to the particle momenta.
As a consequence, $M_{\{\lambda\}}^{(k), \Lambda}$
is constructed from the product of the amplitudes in Eqs.~(\ref{eq:h1},\ref{eq:h2}) and can be expressed as in Eq.~\eqref{eq:proposal}, but with the set of angles $(\alpha,\beta,\gamma) \to (\alpha^k,\beta^k,\gamma^k)$ specific to the aligned configuration having particle $k$ as the spectator.
Since these sets are different, the sum of three amplitudes in Eq.~\eqref{eq:isobar} does not immediately factorize into a product of a single overall rotation function times $O^\nu_{\{ \lambda\}}$.
Fortunately, since the three aligned configurations are defined in the same CM frame, they are related to each other
by a rotation of angle $\hat{\theta}_{k(1)}$ about the $y$ axis (see Eq.~\eqref{eq:rotation.equality}).
Applying such a rotation to bring the configurations with spectator particles $2$ or $3$ to that with particle $1$ as spectator transforms the sum of three amplitudes in Eq.~\eqref{eq:isobar}
into the helicity amplitude $O^\nu_{\{\lambda\}}$ of Eq.~\eqref{eq:proposal},
with $(\alpha,\beta,\gamma) \equiv (\alpha^1,\beta^1,\gamma^1)$.
We shall refer to this aligned configuration corresponding to the spectator particle $1$
(bottom left in Fig.~\ref{fig:Euler.angles}) as the \textit{canonical} configuration.
Finally, we note that before the amplitude in Eq.~\eqref{eq:h2} can be combined with that of Eq.~\eqref{eq:h1},
the former has to be boosted from the isobar rest frame to the space-fixed CM.
Owing to the noncommutativity of Lorentz boosts,
this induces a \textit{Wigner rotation} which affects the helicities of particles $i$ and $j$~\cite{Perl:1974xyz}.
When working with the aligned configurations,
the Wigner rotations are around the $y$ axis and, therefore, are real functions of the Mandelstam variables.
As a result, the final form of the Dalitz-plot function in the canonical configuration is given by
\begin{align} \label{eq:master}
  O^{\nu}_{\{\lambda\}}(\{\sigma\})
  &= \sum_{(ij)k}\sum_{s}^{(ij)\to i,j}\sum_\tau \sum_{\{\lambda'\}}
            n_Jn_s\,
            d^{J}_{\nu,\tau-\lambda_k'}(\hat{\theta}_{k(1)}) \HellH{0\to(ij),k}_{\tau,\lambda_k'}\,
            X_s(\sigma_k)\,
            d^{s}_{\tau,\lambda_i'-\lambda_j'}(\theta_{ij}) \HellH{(ij)\to i,j}_{\lambda_i',\lambda_j'}\\ \nonumber
            & \hspace{2cm} \times
            d^{j_1}_{\lambda_1',\lambda_1}(\wigner^{1}_{k(0)})\,
            d^{j_2}_{\lambda_2',\lambda_2}(\wigner^{2}_{k(0)})\,
            d^{j_3}_{\lambda_3',\lambda_3}(\wigner^{3}_{k(0)}),
\end{align}
with all of the angles given in terms of Mandelstam variables as shown in Appendix~\ref{sec:angle.via.invariants}.
The first sum in Eq.~\eqref{eq:master} is over the three combinations, $(ij)k\in \{(23)1, (31)2, (12)3\}$,
that correspond to the three different decay chains (see Eq.~\eqref{eq:isobar}), with an isobar denoted either by the pair of particles it decays to, $(ij)$,
or the index of the spectator particle $k$.
For every decay chain there are two helicity couplings, $H$,
and the two Wigner $d$-functions in front of them that
describe the orientation of the decay products in the corresponding binary transition.
The argument of the first $d$-function, $\hat{\theta}_{k(1)}$, is measured in the canonical aligned CM.
It corresponds to the polar angle of the isobar $k$, (the direction opposite to $\vec p_k$),
with respect to the $z$ axis (the direction of $-\vec p_1$ in the canonical configuration).
The argument of the second $d$-function, $\theta_{ij}$, is defined in the isobar rest frame, and corresponds to the polar angle of particle $i$ with respect to the direction opposite to the direction of motion of the particle-$0$, \ie
$-\vec p_0$.
Finally, $\wigner^{i}_{k(0)}$ are the polar angles of the Wigner rotations,
computed in the particle $i$ rest frame (see Fig.~\ref{fig:proton.helicity}).
The upper index refers to the particle, the lower index $k$ sets the considered decay chain, and the label $(0)$ reflects the fact that the set of helicities $\{\lambda\}$ is defined in the rest frame of the resonance.
The unprimed helicity indices are defined in the aligned CM while the primed indices correspond to helicities in the isobar rest frame.
We note that for every decay chain one Wigner rotation is trivial, $\wigner^{i}_{i(0)} = 0$, since the boost to the isobar rest frame is in the direction opposite to the spectator momentum (see Eq.~\eqref{eq:tilde.0.k} in Appendix~\ref{sec:angle.via.invariants}).
The main energy dependence of the spin $s$ isobar is given by the $X_s(\{\sigma\})$ function,
which depends on a single Mandelstam variable, \ie the square of the invariant mass of the isobar.
Implementation of the Eq.~\eqref{eq:master} and the code for many practical examples can be found online~\cite{mikha.github.dpd,jpac.github.dpd,jpac.web.dpd}.
We note that helicity couplings have to be defined within a phase convention of the particle helicity states~\cite{Martin:1970xx}.
We used the no-phase convention. Alternatively, in the Jacob-Wick particle-2 phase convention~\cite{Jacob:1959at}, some helicity couplings change sign,
\begin{align}
h^{0\to(ij),k}_{\tau,\lambda_k'} &= \HellH{0\to(ij),k}_{\tau,\lambda_k'} (-1)^{j_k-\lambda_k'}, &
h^{(ij)\to i,j}_{\lambda_i',\lambda_j'} &=  \HellH{(ij)\to i,j}_{\lambda_i',\lambda_j'}(-1)^{j_j-\lambda_j'},
\end{align}
where $h$ are the particle-2-phase-convention helicity couplings.
The latter convention is useful, for example, when the system needs to be symmetrized on particle permutation (identical particles in the final state).
It is often convenient to parametrize the helicity couplings in the $LS$ scheme~\cite{Jacob:1959at}:
\begin{align} \label{eq:LS1}
  \HellH{0\to(ij),k}_{\tau,\lambda_k'}
   &= \sum_{LS}
    \HellH{0\to(ij),k}_{LS}\,
   \sqrt{\frac{2L+1}{2J+1}}\,
        \braket{s,\tau;j_k,-\lambda_k'}{S,\tau-\lambda_k'}\braket{L,0;S,\tau-\lambda_k'}{J,\tau-\lambda_k'},
\end{align}
where $S$ is the spin of the isobar-spectator system and $L$ is the relative orbital angular momentum.
The expressions inside the brackets are the Clebsch-Gordan coefficients.
The other helicity couplings between the isobar and its decay products,
$\HellH{(ij)\to i,j}_{\lambda_i',\lambda_j'}$, are mapped onto the $LS$ couplings through
\begin{align} \label{eq:LS2}
      \HellH{(ij)\to i,j}_{\lambda_i',\lambda_j'}
  &= \sum_{l's'} \HellH{(ij)\to i,j}_{l's'}
        \sqrt{\frac{2l'+1}{2s+1}}
        \braket{j_i,\lambda_i';j_j,-\lambda_j'}{s',\lambda_i'-\lambda_j'}
        \braket{l',0;s',\lambda_i'-\lambda_j'}{s,\lambda_i'-\lambda_j'}.
\end{align}
Parity conservation is straightforward to enforce in the $LS$ scheme since a change in the orbital angular momentum by one unit flips the parity.
Hence, parity conservation makes some $LS$ couplings vanish in the amplitude construction.
The helicity couplings are mass dependent due to the threshold factors~\cite{Collins:1977jy, Martin:1970xx}.
For vanishing breakup momentum $p$, the $LS$ couplings go to zero as $ H_{LS} \propto p^L$.
In Ref.~\cite{Mikhasenko:2017rkh, Pilloni:2018kwm} we showed how this behavior enforces kinematical relations among the helicity amplitudes.
Alternatively, one can use Eqs.~(\ref{eq:LS1},\ref{eq:LS2}) to determine the threshold behavior of the helicity couplings.
The kinematic constraints also exist at pseudothresholds and at the $\sigma=0$ point~\cite{Wang:1966zza, Hara:1964zza, Collins:1977jy, Jackson:1968rfn, Cohen-Tannoudji:1968kvr, Martin:1970xx,Mikhasenko:2017rkh,Pilloni:2018kwm}.
These, however, are typically outside the physical region.\xspace\footnote{For example, the parametrization of dynamic functions suggested in Ref.~\cite{Mikhasenko:2017rkh} for $B\to\psi\pi K$
removes singularities at several unphysical points: $m_{\pi K}^2 = 0$, and $m_{\psi\pi}^2 = 0$,
which are present otherwise when Eq.~\eqref{eq:HFl} and Eq.~\eqref{eq:LSLS} are used.
For the $\Lambda_b^0 \to p\,J/\psi, K^- $ amplitude studied in Ref.~\cite{Pilloni:2018kwm}, the pseudothresholds (out of the physical region as well) also require special consideration.}
A customary form of the $LS$ couplings is
\begin{equation} \label{eq:HFl}
   \HellH{}_{LS} = p^L\, B'_L\,\hellH{}_{LS},
\end{equation}
where $B_{L}'$ are Blatt-Weisskopf factors~\cite{VonHippel:1972fg,Tanabashi:2018oca}, and $\hellH{}_{LS}$ are constant parameters.
The formulation of decay amplitudes in terms of an energy-dependent function $X_s$ times $LS$ couplings is convenient practically.
However, both contribute to the \textit{isobar lineshape}, and they cannot be disentangled in a model-independent way.
The latter reads
\begin{equation} \label{eq:LSLS}
  X_s^{LS;l's'}(\sigma) = \HellH{0\to(ij),k}_{LS} \, X_s \, \HellH{(ij)\to i,j}_{l's'}.
\end{equation}
We note that $X_s^{LS;l's'}$ is the only model-dependent component of Eq.~\eqref{eq:master}.
While the lineshape functions with the same index $s$ need to contain the same set of resonance poles,
they are different for different $LS$, $l's'$, and are unknown from first principles.
Nevertheless, a framework fulfilling unitarity, analyticity, and crossing symmetry, pioneered by Khuri and Treiman~\cite{Khuri:1960zz} (KT),
can be used to calculate the $X_s^{LS;l's'}$ given the two-body elastic scattering phase shift of the relevant subchannels.
The solution of KT equations establishes how the rescattering affects the isobar lineshapes,
which indeed appear to be slightly different in different partial waves ($LS$, $l's'$),
as well as dependent on the mass of particle 0~\cite{Aitchison:1966lpz, Aitchison:1979fj, Niecknig:2012sj, Danilkin:2014cra, Niecknig:2015ija, Mikhasenko:2016mox}.
Equation~\eqref{eq:master} gives a convenient basis for generalization of the KT equations for a system of particles with spin
(see Ref.~\cite{Albaladejo:2019huw} for a complementary method).

Additional constraints arise from isospin symmetry which implies that couplings $\HellH{}_{LS} \to \HellH{}_{LS,I}$ are the same in channels related by rotations in the isospin space,
with the relative strength between individual charge states determined by the Clebsch-Gordan coefficients,
\begin{equation*} %\label{eq:isospin}
  C_{\mu;\mu_i,\mu_j,\mu_k}^{I_{ij}} = \braket{I_i,\mu_i;I_j,\mu_j}{I_{ij},\mu_i+\mu_j} \braket{I_{ij},\mu_i+\mu_j;I_k,\mu_k}{I,\mu}.
\end{equation*}
Here $I_i,\mu_i$ with $i=1,2,3$, and $I,\mu$ are the isospin and its component for the
final-state and decay particles respectively, and  $I_{ij}$ is the total isospin of the $ij$ subsystem. One consequence of isospin symmetry is that $I_{ij} + s'$ (see Eq.~\eqref{eq:LS2}) must be even if particles $i$, and $j$ are identical bosons.

The construction of the decay amplitude presented above can be generalized to some specific cases of more particles in the final state,
in particular, to include subsequent two-body decays, $0\to 1\,2\,3$ with $1\to 4\,5$, which are important for determining the polarization of $1$, \eg in $J/\psi \to \mu^+\mu^-$ or $\Lambda \to pK^-$.
For such decays, the total amplitude can be written as a sum of products of the $0 \to 1\,2\,3$ and the $1\to 4\,5$ amplitudes.
In the canonical configuration, the sum is over the helicity of particle $1$, and the decay amplitude $1 \to 4,5$ is evaluated in the helicity frame for this decay. We illustrate this case in specific examples below.

%
%  _|_|_|_|                                                _|
%  _|        _|    _|    _|_|_|  _|_|_|  _|_|    _|_|_|    _|    _|_|      _|_|_|
%  _|_|_|      _|_|    _|    _|  _|    _|    _|  _|    _|  _|  _|_|_|_|  _|_|
%  _|        _|    _|  _|    _|  _|    _|    _|  _|    _|  _|  _|            _|_|
%  _|_|_|_|  _|    _|    _|_|_|  _|    _|    _|  _|_|_|    _|    _|_|_|  _|_|_|
%                                                _|
%                                                _|

\section{Examples} \label{sec:examples}

\subsection{$\Lambda_c^+\to p K^-\pi^+$ decay chain}  \label{sec:Lc}
$\Lambda_c^+\to p K^-\pi^+$ is the main hadronic decay of the ground-state charmed baryon $\Lambda_c^+$~\cite{Tanabashi:2018oca}.
The measurement of the decay is facilitated by the fact that all final-state particles are charged~\cite{Aaij:2017rin,Yang:2015ytm}.
Each of the three subchannels has at least one clearly visible resonance in the Dalitz plot,
$\Lambda(1520)$ in the $pK^-$ channel, $\Delta(1232)^{++}$ in $p\pi^+$, and $\overline{K}^*(892)^{0}$ in $K^-\pi^+$~\cite{Aitala:1999uq,Yang:2015ytm,Konig:1993wz}.
Furthermore, the decay is supposed to contain a signal of the $\Lambda(1405)$, which might be the manifestation of two different states according to predictions of the Unitarized Chiral Perturbation Theory~\cite{Lambda1405:Tanabashi:2018oca}),
and an intriguing narrow structure seen at the $\Lambda\eta$ threshold in the $pK^-$ invariant mass~\cite{Liu:2019dqc}.
Finally, this decay gives a good handle on the measurement of the $\Lambda_c^+$ polarization, which is important for studying quark hadronization mechanisms~\cite{Falk:1993rf} and
for putting limits on the electric dipole moment which is sensitive to physics beyond the Standard Model~\cite{Botella:2016ksl}.
The amplitude analysis of this decay was performed  in a single study of a small sample of 946 events collected in the E971 experiment~\cite{Aitala:1999uq, Fox:1999ja}. % Endler:2005xz
Given the interest in this reaction and significantly larger data samples gathered by the Belle and LHCb experiments, a new amplitude analysis is called for~\cite{Aaij:2017rin,Yang:2015ytm}.
We are providing a convenient framework for such an analysis. Based on the Dalitz-plot decomposition, Eq.~\eqref{eq:proposal}, the amplitude reads,
\begin{equation} \label{eq:proposal.Lc} %\checkedby{misha}
  \begin{split}
    \includegraphics{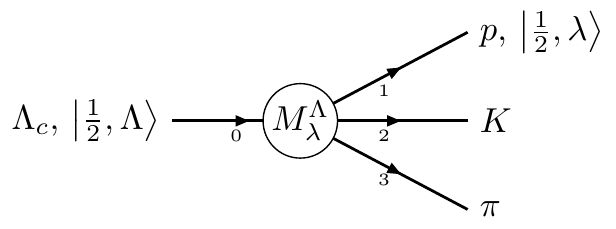}
  \end{split}\qquad
  \begin{split}
    M_{\lambda}^{\Lambda} = \sum_{\nu} D_{\Lambda,\nu}^{1/2*}(\phi_1,\theta_1,\phi_{23}) O_{\lambda}^{\nu}(\{\sigma\}),
  \end{split}
\end{equation}
where $\lambda$ is the proton helicity in the rest frame of $\Lambda_c$.
The Dalitz-plot function $O_{\lambda}^{\nu}(\{\sigma\})$ is given by (\cf Eq.~\eqref{eq:master}),
\begin{align} \label{eq:O.Lc} %\checkedby{misha}
  O_{\lambda}^{\nu}(\{\sigma\})
  &= \sum_{s}^{K^*\to K \pi}\sum_\tau
            \sqrt{2}n_s\,
            \delta_{\nu,\tau-\lambda} \HellH{0\to(23),1}_{\tau,\lambda} \,
            \Props{\sigma_1}{}\,
            d^{s}_{\tau,0}(\theta_{23}) \HellH{(23)\to 2,3}_{0,0}\\ \nonumber
  &\quad + \sum_{s}^{\Delta\to \pi p}\sum_{\tau,\lambda'}
            \sqrt{2}n_s\,
            d^{1/2}_{\nu,\tau}(\hat{\theta}_{2(1)}) \HellH{0\to(31),2}_{\tau,0}\,
            \Props{\sigma_2}{}\,
            d^{s}_{\tau,-\lambda'}(\theta_{31}) \HellH{(31)\to 3,1}_{0,\lambda'}\,
            d^{1/2}_{\lambda',\lambda}(\wigner^{1}_{2(1)})\\ \nonumber
  &\quad + \sum_{s}^{\Lambda\to p K}\sum_{\tau,\lambda'}
            \sqrt{2}n_s\,
            d^{1/2}_{\nu,\tau}(\hat{\theta}_{3(1)}) \HellH{0\to(12),3}_{\tau,0}\,
            \Props{\sigma_3}{}\,
            d^{s}_{\tau,\lambda'}(\theta_{12}) \HellH{(12)\to 1,2}_{\lambda',0}\,
            d^{1/2}_{\lambda',\lambda}(\wigner^{1}_{3(1)}),
\end{align}
where the three lines in Eq.~\eqref{eq:O.Lc} correspond to the three different decay chains, and where we used
$d_{\lambda\lambda}^j(0) = \delta_{\lambda',\lambda'}$ and
$\hat\theta_{1(1)} = 0$, $\wigner_{1(0)}^1 = \wigner_{1(1)}^1= 0$ (see Eq.~\eqref{eq:tilde.0.k}) for the first decay chain.
We also replaced $\wigner_{2(0)}^1$ and $\wigner_{3(0)}^1$ with $\wigner_{2(1)}^1$ and $\wigner_{3(1)}^1$ in the second and third chains, respectively (see Appendix~\ref{sec:angle.via.invariants}).

For studies of $\Lambda_c$ polarization, the decay amplitude needs to be contracted with the polarization matrix as given in Eq.~\eqref{eq:decay}.
For a spin-$1/2$ particle, $\rho = (\mathds{1}+ \vec P \cdot \vec \sigma)$ with $\vec P$ being the polarization vector, and $\vec \sigma$ the Pauli matrices.
By choosing the $z$ axis of the space-fixed CM in the direction of polarization,\footnote{For example, if the $\Lambda_c$ is produced by parity-conserving interactions,
the polarization must be perpendicular to the production plane.} the expression for the cross section reads,
\begin{align} \label{eq:Lc.polarization}
    \frac{\diff \sigma}{\diff \cos\theta_{1} \diff\phi_{23} \,\diff \sigma_1 \diff \sigma_2} &= N_0
    \sum_{\nu\nu'}
    \begin{pmatrix}
    1+P\cos\theta_{1} & -P \sin\theta_{1}\,e^{i\phi_{23}}\\
    -P \sin\theta_{1}\,e^{-i\phi_{23}} & 1-P\cos\theta_{1}
    \end{pmatrix}_{\nu\nu'}
    \sum_{\lambda} O_{\lambda}^{\nu} O_{\lambda}^{\nu'*},
\end{align}
with $N_0$ being a normalization constant, and $P = |\vec P|$. The angles of the first decay chain ($\theta_1$, $\phi_{23}$) are used in the polarization matrix
in agreement with Eq.~\eqref{eq:O.Lc}.

One finds that Eq.~\eqref{eq:O.Lc} differs from the model used in Ref.~\cite{Aitala:1999uq} due to the presence of the Wigner rotations
(the $\wigner$ angles in the second and the third decay chains do not appear in Tables~3 and~4 of~\cite{Aitala:1999uq}).
As discussed above, these rotations are required for a consistent description of the proton helicity states.
In addition, the model of Ref.~\cite{Aitala:1999uq} does not permit a decomposition as in Eq.~\eqref{eq:proposal}
and results in an unphysical dependence on $\phi_{23}$, even for unpolarized $\Lambda_c$.

\subsection{$\overline{B}^0\to \psi\pi^+ K^-$ decay chain}

Amplitude analysis of the $\overline{B}^0\to \psi(2S)\pi^+ K^-$ decay was performed by Belle~\cite{Mizuk:2009da,Chilikin:2013tch} and LHCb~\cite{Aaij:2014jqa,Aaij:2015zxa,Aaij:2019ipm}
revealing the exotic-charmonium candidate $Z_c(4430)^+$~\cite{Esposito:2016noz,Olsen:2017bmm}.
The signal is also seen in $\overline{B}^0\to J/\psi \pi^+ K^-$, where hints of other exotic structures also appear~\cite{Aaij:2019ipm,Chilikin:2014bkk}.
In the first analysis by Belle only the Dalitz-plot distribution was fitted~\cite{Mizuk:2009da}.
In subsequent analyses, the angular distribution of the muon pairs from the $\psi(2S)$ decays was included~\cite{Chilikin:2013tch,Aaij:2014jqa}.
Although the amplitudes used in these analyses are consistent with each other and with our method (see Appendix~\ref{sec:dec.B}),
we believe that our formulation is more transparent.
The amplitude for the decay chain $\overline{B}^0\to \mu^+\mu^- \pi^+ K^-$,
can be split into two parts $\overline{B}^0\to \psi \pi^+ K^-$ and $\psi \to \mu^+ \mu^-$, denoted $A$ and $B$, respectively (see the diagram below).
The angular dependence is factored out according to Eq.~\eqref{eq:proposal} for both decays:
\begin{equation}\label{eq:DPD} %\checkedby{misha}
  \begin{split}
    \includegraphics{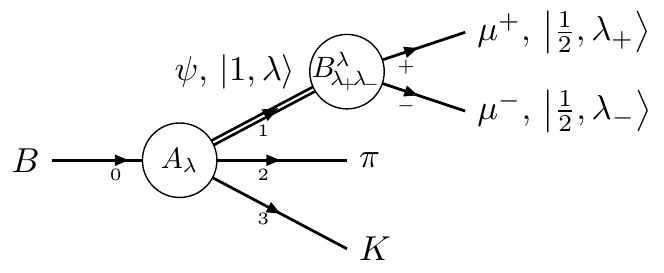}
  \end{split}\qquad
  \begin{split}
    M_{\lambda_+,\lambda_-} &= \sum_\lambda A_{\lambda}\,  B^{\lambda}_{\lambda_+\lambda_-}\\
    &=
    \sum_{\lambda} O_{\lambda}(\{\sigma\}) \left[\sqrt{3} D^{1*}_{\lambda,\lambda_+-\lambda_-}(\phi_+,\theta_+,0) \HellH{1\to \mu^+\!,\mu^-}_{\lambda_+,\lambda_-} \right],
  \end{split}
\end{equation}
with $\lambda$ being the helicity of $J/\psi$ in the space-fixed CM.
The muon helicities $\lambda_+$ and $\lambda_-$ are defined in the $J/\psi$ rest frame obtained by a boost against the $p_1$ momentum from the canonical confirmation.
We note that when a different frame is used to define the muon helicities the Wigner rotations for muon states might appear, which, however, cancel out in the expression for the cross section if muon helicities are summed over.
The overall $D$ function that rotates the canonical configuration to the actual one is absent because the $B$ has spin zero.
For the $\psi$ decay amplitude, the spherical angles $(\phi_+,\theta_+)$ are the angles of $\mu^+$ in the $\psi$ helicity frame,
reached from the aligned CM by a boost in direction of $-\vec p_1$.
Hence, the azimuthal angle $\phi_+$ is
equal to the angle between the $B$ meson decay plane and the plane containing the muon pair in the $B$ rest frame.
As customary, the helicity amplitude $\HellH{1\to \mu^+\!,\mu^-}_{\pm 1/2,\mp 1/2}$ can be neglected since $m_\psi \gg m_\mu$.
The Dalitz-plot function is given by
\begin{align} \label{eq:O.Bdecay} %\checkedby{misha}
  O_{\lambda}(\{\sigma\})
  &= \sum_{s}^{K^*\to K\pi}
            n_s\,
            \HellH{0\to(23),1}_{\lambda,\lambda}\,
            \Props{\sigma_1}{}\,
            d^{s}_{\lambda,0}(\theta_{23}) \HellH{(23)\to 2,3}_{0,0}\\ \nonumber
  &\quad + \sum_{s}^{Z\to\psi \pi}\sum_{\lambda'}
            n_s\,
            \HellH{0\to(12),3}_{0,0}\,
            \Props{\sigma_3}{}\,
            d^{s}_{0,\lambda'}(\theta_{12}) \HellH{(12)\to 1,2}_{\lambda',0}\, d^{1}_{\lambda',\lambda}(\wigner^{1}_{3(1)}).
\end{align}
The Wigner rotation on the second line
appears because the $J/\psi$, which in the $Z$-isobar chain has the spin quantized along the $\pi$ direction, is boosted from the $Z$ rest frame to the $B$ rest frame.
Equation~\eqref{eq:O.Bdecay} is equivalent to the amplitude used in the two-dimentional analysis of Ref.~\cite{Mizuk:2009da}.
The extension to the a four-dimensional analysis that includes the muon angular distribution is as simple as Eq.~\eqref{eq:DPD},
and its equivalence with the method used in~\cite{Chilikin:2013tch} is demonstrated in Appendix~\ref{sec:dec.B}.

\subsection{$\Lambda_b^0\to p\,K^-\,J/\psi$ decay chain}

Pentaquark candidates were discovered in the reaction $\Lambda_b^0\to p\,K^-\,J/\psi[\to\mu^+\mu^-]$ as peaks in the $J/\psi p$ invariant mass distribution~\cite{Aaij:2015tga,Aaij:2016phn,Aaij:2019vzc}.
The amplitude analysis of Ref.~\cite{Aaij:2015tga} covers the full six-dimensional phase space distribution:
two of the three Euler angles that determine the orientation of the $\Lambda_b^0$  decay plane,
the two Dalitz-plot variables, and the two angles which determine the distribution of the muon pair from $J/\psi$ decay.
Both decay chains with $\Lambda^0/\Sigma^0$ isobars in the $pK^-$ subchannel and $P_c$ isobars in $J/\psi p$ subchannel
are described as a product of the amplitudes in Eq.~\eqref{eq:h1} and Eq.~\eqref{eq:h2}, and of the $J/\psi \to \mu^+\mu^-$ amplitude.
The muon angles are measured in the $J/\psi$ rest frame obtained by a boost from the isobar rest frame in each decay chain.
It was realized that these two different $J/\psi$ helicity frames differ only by an azimuthal rotation that is compensated for when the two decay chains are summed up.
The Wigner rotation for the proton state was found to be a rotation about $y$ and therefore to be real.

In our construction, we factorize the $J/\psi$ decay analogously to Eq.~\eqref{eq:DPD}.
The Euler angles for the decay-plane orientation appear for both the $1\to 3$ decay of $\Lambda_b^0$ and the $J/\psi\to \mu^+\mu^-$ decay.
\begin{equation} \label{eq:Lb.rot} %\checkedby{misha}
  \begin{split}
      \includegraphics[scale=0.9]{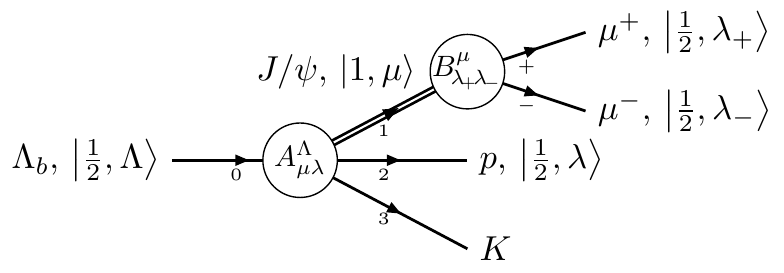}
  \end{split}\qquad
  \begin{split}
    M^{\Lambda}_{\lambda;\lambda_+\lambda_-} &=
    \sum_\mu A^{\Lambda}_{\mu\lambda} B^{\mu}_{\lambda_+\lambda_-}\\
    &= \sum_{\nu\mu} D_{\Lambda,\nu}^{1/2}(\phi_1,\theta_1,\phi_{23}) O^{\nu}_{\lambda\mu}(\{\sigma\})\\
    &\qquad\qquad \times \left[
    \sqrt{3} D^{1}_{\mu,\lambda_+-\lambda_-}(\phi_+,\theta_+,0) \HellH{1\to \mu^+\!,\mu^-}_{\lambda_+,\lambda_-}\right],
  \end{split}
\end{equation}
with the term in the brackets describing the decay $J/\psi \to \mu^+\mu^-$.
The isobar decomposition of the Dalitz-plot function for $\Lambda_b^0\to J/\psi p K^-$ gives,
\begin{align} \label{eq:O.Ldecay} %\checkedby{misha}
  O^{\nu}_{\lambda\mu}(\{\sigma\})
  &= \sum_{s}^{\Lambda,\Sigma\to p K}\sum_\tau
            \sqrt{2}n_s\,
            \delta_{\nu,\tau-\mu} \HellH{0\to(23),1}_{\tau,\mu}\,
            \Props{\sigma_1}{}\,
            d^{s}_{\tau,\lambda'}(\theta_{23}) \HellH{(23)\to 2,3}_{\lambda',0}\,
            d^{1/2}_{\lambda',\lambda}(\wigner_{1(2)}^2) \\ \nonumber
  &\quad + \sum_{s}^{P_c\to J/\psi p}\sum_{\tau,\mu',\lambda'}
           \sqrt{2}n_s\,
            d^{1/2}_{\nu,\tau}(\hat{\theta}_{3(1)}) \HellH{0\to(12),3}_{\tau,0}\,
            \Props{\sigma_3}{}\,
            d^{s}_{\tau,\mu'-\lambda'}(\theta_{12}) \HellH{(12)\to 1,2}_{\mu',\lambda'}\,
            d^{1}_{\mu',\mu}(\wigner^{1}_{3(1)})\,
            d^{1/2}_{\lambda',\lambda}(\wigner^{2}_{3(2)}),
\end{align}
where $\sigma_{1} = m_{p K}^2$, and $\sigma_3 = m_{J/\psi p}^2$.
In the $0\to 1\,2\,3$ decay, there are two particles with spin in the final state, $J/\psi$ and the proton.
In chain-$1$, which contains the hyperons, $J/\psi$ (particle-$1$) is the spectator and the Wigner rotation
applies to the proton only (particle-$2$), which is boosted from the hyperon rest frame to the $\Lambda_b$ rest frame.
The second line of Eq.~\eqref{eq:O.Ldecay} provides the amplitude for the $P_c$ decay chain, (chain-$3$)
in which both $J/\psi$ and proton are boosted from the $P_c$ rest frame to the $\Lambda_b$ rest frame and thus are both affected by a Wigner rotation.
As above, the helicity amplitude $\HellH{1\to \mu^+\!,\mu^-}_{\pm 1/2,\mp 1/2}$ can be neglected since $m_\psi \gg m_\mu$.
The cross section for polarized $\Lambda_b$ can be constructed analogously to Eq.~\eqref{eq:Lc.polarization}.

\section{Conclusions}
Modern hadron spectroscopy and beyond the standard model searches often rely on amplitude analyses of multibody decays.
The treatment of such decays necessitates the construction of multidimensional models able to separate the contributions of the various physical processes.
However, the conventional way to build amplitudes mixes up angular variables (which give the orientation of the decay plane and provide information about the polarization of the decaying particle),
and the dynamical variables such as the invariant masses of the decay subsystems (which provide information about the intermediate resonances).

We have proposed an amplitude construction that separates the angular variables from the dynamical variables in a model-independent way.
For the $0\to 1\,2\,3$ transition we have built a formalism that factors out the decay-plane orientation in such a way that the remaining dynamical function depends only on two invariant quantities,
as required by the general principles.
This dynamical function, the Dalitz-plot function, is subject to modeling.
All angles required by the isobar model construction are known functions of invariant variables.
The calculation of the angles in our approach does not require boosts or rotations between different frames, simplifying numerical calculations relative to the other approaches.
Moreover by explicitly aligning particles in the decay plane, all rotations appearing in Eq.~\eqref{eq:master} are real functions.
Therefore the phases arising in the amplitudes, besides the overall rotation in Eq.~\eqref{eq:proposal}, are caused by dynamical reasons only.

In the formalism we proposed in this work, it is straightforward to maintain the consistency of the helicity states between different decay channels as enforced by Lorentz invariance.
The remaining dynamical information, that, for example, distinguishes the tensor approach from the helicity formalism, appears in the isobar lineshape functions only.
The latter are model dependent, and the differences between different models can be taken as theoretical uncertainties.

The amplitude formulations used by Belle and LHCb to analyze the $\overline{B}^0\to \psi\pi^+ K^-$ and $\Lambda_b^0\to J/\psi\,p\,K^-$ followed by $\psi (J/\psi) \to\mu^+\mu^-$~\cite{Chilikin:2013tch,Aaij:2015tga}
produce the same matrix element as our formulation.
This illustrates that coherence between different two-body decay chains in three-body decay can be achieved either by aligning
helicity states of the final-state particles (here $\mu^+\mu^-$) or helicity states of the long-lived factorizable intermediate particle (here $\psi$).
However, the latter approach, which we advocate, produces simpler formulae, which are not only faster to evaluate, but also explicitly reveal factorization of the matrix element
into the part describing probability density on the Dalitz plane, and parts describing decay angles of any possible quasi-stable particles (here $B$ and $\Lambda_b^0$ as well as $\psi$).
Such factorization holds, but it is not obvious from the formulae in the former approach.
The approach that we have proposed will also make it easy for experimentalists to share the code between 2D Dalitz-plot analyses and extensions of the amplitude fits to more decay dimensions.
The framework is being actively tested in several LHCb analyses, the code and more practical information can be found in~\cite{mikha.github.dpd,jpac.github.dpd,jpac.web.dpd}.

\begin{acknowledgments}
We thank Daniele Marangotto and Anton Poluektov for several motivating discussions on the issue of $\Lambda_c^+$ anisotropy.
The first ideas on this project were presented at the HPSS (Hadron Physics Summer School) in J\"ulich,
and we would like to thank Sebastian Neubert, the students joining the working group, and organizers of the school.
This work is supported by the U.S.~Department of Energy Grants No.~DE-AC05-06OR23177 and No.~DE-FG02-87ER40365,
the U.S.~National Science Foundation under Grant No.~PHY-1415459, % Adam, PIF
No.~PHY-1803004, % Tomasz
by PAPIIT-DGAPA (UNAM, Mexico) Grant No.~IA101819, % Cesar
and by CONACYT (Mexico) Grants No.~251817 and~No.~A1-S-21389, % Cesar
and by Polish Science Center (NCN) Grant No.~2018/29/B/ST2/02576. % Lukasz
We acknowledge support from STFC (United Kingdom).  % marco & Sara
V.M. is supported by Comunidad Aut\'onoma de Madrid through
Programa de Atracci\'on de Talento Investigador 2018 (Modalidad 1).
\end{acknowledgments}

\pagebreak

%
%                                                          _|  _|
%    _|_|_|  _|_|_|    _|_|_|      _|_|    _|_|_|      _|_|_|      _|    _|
%  _|    _|  _|    _|  _|    _|  _|_|_|_|  _|    _|  _|    _|  _|    _|_|
%  _|    _|  _|    _|  _|    _|  _|        _|    _|  _|    _|  _|  _|    _|
%    _|_|_|  _|_|_|    _|_|_|      _|_|_|  _|    _|    _|_|_|  _|  _|    _|
%            _|        _|
%            _|        _|

\appendix

\section{Expression for the angles in Dalitz-plot representation}
\label{sec:angle.via.invariants}

The isobar model construction for a general $0\to 1\, 2\, 3$ decay shown in Fig.~\ref{fig:general.1to3} contains
multiple polar angles, which either are used to specify the direction of a final-state particle in a specific frame
($\hat\theta_{k}$ and $\theta_{ij}$),
or appear to account for the change of a helicity state upon boosts ($\wigner_{i(j)}^{k}$).
The cosine of these angles can be explicitly expressed in terms of invariant variables.
All of the angles discussed above are polar, defined in the range $[0,\pi]$, which makes their determination as
a function of the cosine unique.

The scattering angle $\theta_{ij}$ is defined in the rest frame of the isobar in the $(ij)$ channel, and it is the relative angle between particle $i$ and the spectator particle $k$  (see Fig.~\ref{fig:proton.helicity}).
Explicitly,
\begin{align}%\checkedby{misha}
  \cos(\theta_{12}) &= \frac{2\sigma_3(\sigma_2-m_3^2-m_1^2)-(\sigma_3+m_1^2-m_2^2)(m_0^2-\sigma_3-m_3^2)}{\lambda^{1/2}(m_0^2,m_3^2,\sigma_3)\lambda^{1/2}(\sigma_3,m_1^2,m_2^2)},\quad\tikznode{ps3}{}&&\quad\\\nonumber
  \cos(\theta_{23}) &= \frac{2\sigma_1(\sigma_3-m_1^2-m_2^2)-(\sigma_1+m_2^2-m_3^2)(m_0^2-\sigma_1-m_1^2)}{\lambda^{1/2}(m_0^2,m_1^2,\sigma_1)\lambda^{1/2}(\sigma_1,m_2^2,m_3^2)},\quad\tikznode{ps1}{}&&\quad\\\nonumber
  \cos(\theta_{31}) &= \frac{2\sigma_2(\sigma_1-m_2^2-m_3^2)-(\sigma_2+m_3^2-m_1^2)(m_0^2-\sigma_2-m_2^2)}{\lambda^{1/2}(m_0^2,m_2^2,\sigma_2)\lambda^{1/2}(\sigma_2,m_3^2,m_1^2)}.\quad\tikznode{ps2}{}&&\quad
\end{align}
Arrows on the side of the equation show how the indices are related by cyclic permutations.
\begin{tikzpicture}[remember picture,overlay,black]
  \draw[<->,black] (ps3) -- +(0.5,0) |- node[xshift=2mm,right,rotate=90] {\scriptsize $(123)$} (ps1);
  \draw[<->,black] (ps3) -- +(1.2,0) |- node[xshift=2mm,right,rotate=90] {\scriptsize $(321)$} (ps2);
\end{tikzpicture}

The angle $\hat\theta_{k(i)}$ gives the direction of the isobar in the chain-$k$ given the canonical chain-$i$ used for the alignment.
Throughout the paper the canonical chain corresponds to $i=1$, thus only $\hat\theta_{k(1)}$ are needed.
In general, $\hat\theta_{k(i)}$ is defined in the aligned CM frame as the angle between the direction of isobar $k$ and the direction opposite to particle $i$, so that
\begin{equation}
  \hat\theta_{1(1)} = \hat\theta_{2(2)} = \hat\theta_{3(3)} = 0.
\end{equation}
For the angles with sequential index order, one finds
\begin{align}
  \cos(\hat\theta_{3(1)}) &= \frac{(m_0^2+m_3^2-\sigma_3)(m_0^2+m_1^2-\sigma_1) - 2m_0^2(\sigma_2-m_3^2-m_1^2)}{\lambda^{1/2}(m_0^2,m_1^2,\sigma_1)\lambda^{1/2}(m_0^2,\sigma_3,m_3^2)},\quad\tikznode{ph1}{}&&\quad\\\nonumber
  %\checkedby{misha,vincent}
  \cos(\hat\theta_{1(2)}) &= \frac{(m_0^2+m_1^2-\sigma_1)(m_0^2+m_2^2-\sigma_2) - 2m_0^2(\sigma_3-m_1^2-m_2^2)}{\lambda^{1/2}(m_0^2,m_2^2,\sigma_2)\lambda^{1/2}(m_0^2,\sigma_1,m_1^2)},\quad\tikznode{ph2}{}&&\quad\\\nonumber
  \cos(\hat\theta_{2(3)}) &= \frac{(m_0^2+m_2^2-\sigma_2)(m_0^2+m_3^2-\sigma_3) - 2m_0^2(\sigma_1-m_2^2-m_3^2)}{\lambda^{1/2}(m_0^2,m_3^2,\sigma_3)\lambda^{1/2}(m_0^2,\sigma_2,m_2^2)}.\quad\tikznode{ph3}{}&&\quad
\end{align}
\begin{tikzpicture}[remember picture,overlay,black]
  \draw[<->,black] (ph1) -- +(0.5,0) |- node[xshift=2mm,right,rotate=90] {\scriptsize $(123)$} (ph2);
  \draw[<->,black] (ph1) -- +(1.2,0) |- node[xshift=2mm,right,rotate=90] {\scriptsize $(321)$} (ph3);
\end{tikzpicture}
Angles the other order of indices, \eg $\hat\theta_{2(1)}$, imply a clockwise rotation (see Fig.~\ref{fig:proton.helicity}),
which can be realized using rotation about $z$ by $\pi$ before and after:
\begin{equation} \label{eq:R.clockwise}
  R(0,\hat\theta_{2(1)},0) = R(\pi,\hat\theta_{1(2)},\pi),
\end{equation}
in the convention of the Wigner function in Eq.~\eqref{eq:wigner.funciton}.
It results in an extra phase factor,\xspace\footnote{
The clockwise rotation can be also seen as a counterclockwise rotation by a negative angle,
$\hat\theta_{2(1)} = -\hat\theta_{1(2)}$.
The same results is obtained by using the property of the Wigner $d$-function,
$d_{\lambda\lambda'}^J(-\theta) = (-1)^{\lambda-\lambda'} d_{\lambda\lambda'}^J(\theta)$.
}
\begin{align}
  d_{\lambda\lambda'}^{j}(\hat\theta_{2(1)}) &= (-1)^{\lambda-\lambda'} d_{\lambda\lambda'}^{j}(\hat\theta_{1(2)}),\\\nonumber
  d_{\lambda\lambda'}^{j}(\hat\theta_{3(2)}) &= (-1)^{\lambda-\lambda'} d_{\lambda\lambda'}^{j}(\hat\theta_{2(3)}),\\\nonumber
  d_{\lambda\lambda'}^{j}(\hat\theta_{1(3)}) &= (-1)^{\lambda-\lambda'} d_{\lambda\lambda'}^{j}(\hat\theta_{3(1)}).
\end{align}

Equation~\eqref{eq:master} contains nine angles for the Wigner rotation denoted by $\wigner_{k(0)}^i$,
where the upper index specifies which particle is boosted, the lower index $k$ shows which decay chain is aligned,
and number in parentheses indicates the frame where all helicities are defined ($0$ is for the aligned CM;
the nonzero number would correspond to the isobar rest frame in the respective decay chain).
The angle $\wigner_{k(0)}^i$ is equal to the angle between isobar $i$ and isobar $k$ in particle-$i$ rest frame.
The relevant angles can be found using the following relations:
\begin{align} \label{eq:tilde.0.k}
  \wigner^{i}_{k(0)} &= \wigner^{i}_{k(i)} & \wigner_{k(k)}^{i} &= 0, & \forall\,\, k, i &\in \{1,2,3\}.
\end{align}
\mbox{\parbox{\textwidth}{
\begin{align}
    \cos(\wigner_{1(3)}^{1}) &= \frac{2m_1^2(\sigma_2-m_0^2-m_2^2)+(m_0^2+m_1^2-\sigma_1)(\sigma_3-m_1^2-m_2^2)}{\lambda^{1/2}(m_0^2,m_1^2,\sigma_1)\lambda^{1/2}(\sigma_3,m_1^2,m_2^2)},\quad\tikznode{pp1}{}&&\quad\\\nonumber
    \cos(\wigner_{2(1)}^{1}) &= \frac{2m_1^2(\sigma_3-m_0^2-m_3^2)+(m_0^2+m_1^2-\sigma_1)(\sigma_2-m_1^2-m_3^2)}{\lambda^{1/2}(m_0^2,m_1^2,\sigma_1)\lambda^{1/2}(\sigma_2,m_1^2,m_3^2)},\quad\tikznode{pp2}{}&&\quad\\\nonumber
    \cos(\wigner_{2(1)}^{2}) &= \frac{2m_2^2(\sigma_3-m_0^2-m_3^2)+(m_0^2+m_2^2-\sigma_2)(\sigma_1-m_2^2-m_3^2)}{\lambda^{1/2}(m_0^2,m_2^2,\sigma_2)\lambda^{1/2}(\sigma_1,m_2^2,m_3^2)},\quad\tikznode{pp3}{}&&\quad\\\nonumber
    \cos(\wigner_{3(2)}^{2}) &= \frac{2m_2^2(\sigma_1-m_0^2-m_1^2)+(m_0^2+m_2^2-\sigma_2)(\sigma_3-m_2^2-m_1^2)}{\lambda^{1/2}(m_0^2,m_2^2,\sigma_2)\lambda^{1/2}(\sigma_3,m_2^2,m_1^2)},\quad\tikznode{pp4}{}&&\quad\\\nonumber
    \cos(\wigner_{3(2)}^{3}) &= \frac{2m_3^2(\sigma_1-m_0^2-m_1^2)+(m_0^2+m_3^2-\sigma_3)(\sigma_2-m_3^2-m_1^2)}{\lambda^{1/2}(m_0^2,m_3^2,\sigma_3)\lambda^{1/2}(\sigma_2,m_3^2,m_1^2)},\quad\tikznode{pp5}{}&&\quad\\\nonumber
    \cos(\wigner_{1(3)}^{3}) &= \frac{2m_3^2(\sigma_2-m_0^2-m_2^2)+(m_0^2+m_3^2-\sigma_3)(\sigma_1-m_3^2-m_2^2)}{\lambda^{1/2}(m_0^2,m_3^2,\sigma_3)\lambda^{1/2}(\sigma_1,m_3^2,m_2^2)},\quad\tikznode{pp6}{}&&\quad
\end{align}
}}
For the other order of indices the clockwise rotation is implied.
It results in a phase factor as discussed above,
\begin{align} \label{eq:flip.the.sign}
  d_{\lambda\lambda'}^{j}(\wigner^{1}_{1(2)}) &= (-1)^{\lambda-\lambda'} d_{\lambda\lambda'}^{j}(\wigner^{1}_{2(1)}), \\\nonumber
  d_{\lambda\lambda'}^{j}(\wigner^{2}_{2(3)}) &= (-1)^{\lambda-\lambda'} d_{\lambda\lambda'}^{j}(\wigner^{2}_{3(2)}), \\\nonumber
  d_{\lambda\lambda'}^{j}(\wigner^{3}_{3(1)}) &= (-1)^{\lambda-\lambda'} d_{\lambda\lambda'}^{j}(\wigner^{3}_{1(3)}),
\end{align}
for all $k = 1,2,3$.
\begin{tikzpicture}[remember picture,overlay,black]
  \draw[<->,black] (pp1) -- +(0.5,0) |- node[xshift=2mm,right,rotate=90] {\scriptsize $(23)$} (pp2);
  \draw[<->,black] (pp1) -- +(1.2,0) |- node[xshift=2mm,right,rotate=90] {\scriptsize $(123)$} (pp3);
  \draw[<->,black] (pp3) -- +(0.5,0) |- node[xshift=2mm,right,rotate=90] {\scriptsize $(31)$} (pp4);
  \draw[<->,black] (pp1) -- +(2.0,0) |- node[xshift=2mm,right,rotate=90] {\scriptsize $(321)$} (pp5);
  \draw[<->,black] (pp5) -- +(0.5,0) |- node[xshift=2mm,right,rotate=90] {\scriptsize $(12)$} (pp6);
\end{tikzpicture}

The Wigner angles with all different indices (\eg $\wigner_{2(3)}^{1}$) do not enter Eq.~\eqref{eq:master}.
Nevertheless, they are useful in checking numerical implementation. One finds simple sum rules (see Fig.~\ref{fig:proton.helicity}):
\begin{align} \label{eq:wigner.123}
  \wigner^{(1)}_{2(3)} &= \wigner^{(1)}_{2(1)} + \wigner^{(1)}_{1(3)}, &
  \wigner^{(2)}_{3(1)} &= \wigner^{(2)}_{3(2)} + \wigner^{(2)}_{2(1)}, &
  \wigner^{(3)}_{1(2)} &= \wigner^{(3)}_{1(3)} + \wigner^{(3)}_{3(2)},
\end{align}
where
\begin{align}
  \cos(\wigner_{2(3)}^{1}) &= \frac{2m_1^2(m_2^2+m_3^2-\sigma_1)+(\sigma_2-m_1^2-m_3^2)(\sigma_3-m_1^2-m_2^2)}{\lambda^{1/2}(\sigma_2,m_3^2,m_1^2)\lambda^{1/2}(\sigma_3,m_1^2,m_2^2)},\quad\tikznode{pd1}{}&&\quad\\\nonumber
  \cos(\wigner_{3(1)}^{2}) &= \frac{2m_2^2(m_3^2+m_1^2-\sigma_2)+(\sigma_3-m_2^2-m_1^2)(\sigma_1-m_2^2-m_3^2)}{\lambda^{1/2}(\sigma_3,m_1^2,m_2^2)\lambda^{1/2}(\sigma_1,m_2^2,m_3^2)},\quad\tikznode{pd2}{}&&\quad\\\nonumber
  \cos(\wigner_{1(2)}^{3}) &= \frac{2m_3^2(m_1^2+m_2^2-\sigma_3)+(\sigma_1-m_3^2-m_2^2)(\sigma_2-m_3^2-m_1^2)}{\lambda^{1/2}(\sigma_1,m_2^2,m_3^2)\lambda^{1/2}(\sigma_2,m_3^2,m_1^2)}.\quad\tikznode{pd3}{}&&\quad
\end{align}
\begin{tikzpicture}[remember picture,overlay,black]
  \draw[<->,black] (pd1) -- +(0.5,0) |- node[xshift=2mm,right,rotate=90] {\scriptsize $(123)$} (pd2);
  \draw[<->,black] (pd1) -- +(1.2,0) |- node[xshift=2mm,right,rotate=90] {\scriptsize $(321)$} (pd3);
\end{tikzpicture}
The prescription to change the order of the lower is the same as in Eq.~\eqref{eq:flip.the.sign}:
\begin{align}
  d_{\lambda\lambda'}^{j}(\wigner^{1}_{3(2)}) &= (-1)^{\lambda-\lambda'} d_{\lambda\lambda'}^{j}(\wigner^{1}_{2(3)}), \\\nonumber
  d_{\lambda\lambda'}^{j}(\wigner^{2}_{1(3)}) &= (-1)^{\lambda-\lambda'} d_{\lambda\lambda'}^{j}(\wigner^{2}_{3(1)}), \\\nonumber
  d_{\lambda\lambda'}^{j}(\wigner^{3}_{2(1)}) &= (-1)^{\lambda-\lambda'} d_{\lambda\lambda'}^{j}(\wigner^{3}_{1(2)}).
\end{align}

%
%        _|                                                                              _|    _|      _|
%    _|_|_|    _|_|      _|_|_|    _|_|    _|_|_|  _|_|    _|_|_|      _|_|      _|_|_|      _|_|_|_|        _|_|    _|_|_|
%  _|    _|  _|_|_|_|  _|        _|    _|  _|    _|    _|  _|    _|  _|    _|  _|_|      _|    _|      _|  _|    _|  _|    _|
%  _|    _|  _|        _|        _|    _|  _|    _|    _|  _|    _|  _|    _|      _|_|  _|    _|      _|  _|    _|  _|    _|
%    _|_|_|    _|_|_|    _|_|_|    _|_|    _|    _|    _|  _|_|_|      _|_|    _|_|_|    _|      _|_|  _|    _|_|    _|    _|
%                                                          _|
%                                                          _|

\section{Relation to the Belle analyses of $\overline{B}^0\to \psi \pi^+ K^-$} \label{sec:dec.B}

The decay amplitude in our approach is presented in Eq.~\eqref{eq:DPD} and Eq.~\eqref{eq:O.Bdecay}.
However, the amplitudes in Ref.~\cite{Chilikin:2013tch,Aaij:2015tga} are written differently.
The decay of $\psi$ is not separated from the three-body decay of $B$,
but it is taken into account for either decay chain separately by boosting to
the dimuon rest frame from different frames, and defining the corresponding angles accordingly.

The amplitude was constructed using an isobar model with two chains, $K^*$ states in the $\pi K$ subchannel (chain-$3$ in discussion below), and the $Z$ chain (chain-$1$).
Using the notations of this paper the expression for the Belle matrix element reads:
\begin{align} \label{eq:Chilinkin}
  \left(A_{\xi}\right)_\text{Belle} &= \sum_{s,\lambda}\bigg(
  n_s\,
  \HellH{0\to(23),1}_{\lambda,\lambda}\,\,
  X_s(\sigma_1)\,\,
  d_{\lambda,0}^{s}(\theta_{23}) \,\HellH{(23)\to 2,3}_{0,0}\,\,
  D_{\lambda,\xi}^{1*}(\phi_+, \theta_+ )\,\HellH{1\to \mu^+\!,\mu^-}_{\lambda_+,\lambda_-}\\ \nonumber
  & \qquad +
  n_s\,
  \HellH{0\to(12),3}_{0,0}\,\,
  X_s(\sigma_3)\,\,
  d_{0,\lambda}^{s}(\theta_{12})\,\HellH{(12)\to 1,2}_{\lambda,0}\,\,
  D_{\lambda,\xi}^{1*}(\phi_+',\theta_+') \,
  \HellH{1\to \mu^+\!,\mu^-}_{\lambda_+,\lambda_-}\,e^{i\xi\alpha}
  \bigg),
\end{align}
where $(\theta,\phi)$ are spherical angles of $\mu^+$ in the $\psi$ rest frame after the boost from the aligned CM,
while $(\theta',\phi')$ are the spherical angles of $\mu^+$ in the $\psi$ rest frame after the boost from the $(\psi \pi)$ rest frame.
The factor $\exp(i\xi\alpha)$ is added to align the helicities of chain-$3$ with the ones of chain-$1$.
The angle $\alpha$ is defined as the difference of azimuthal angles of $\pi$ and $K^*$ (sum of the vectors of $K$ and $\pi$) in the $\psi$ rest frame~\cite{Chilikin:2013tch}.

\begin{figure}[t]
  \includegraphics[width=0.9\textwidth]{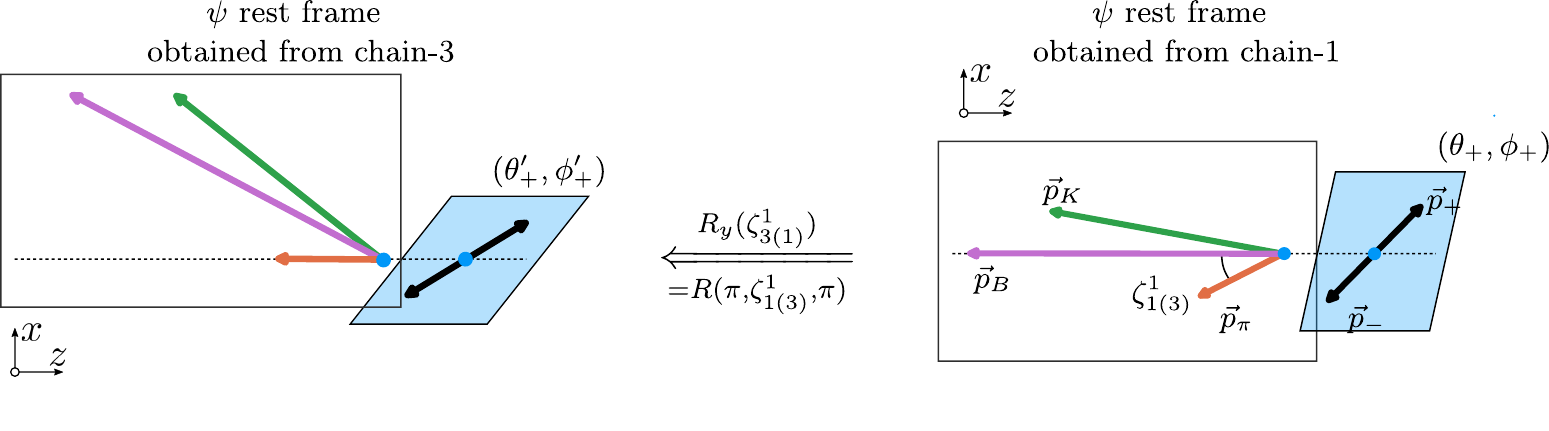}
  \caption {Visual representation for Eq.~\eqref{eq:rotation.equality}.
  }
  \label{fig:alpha.angle}
\end{figure}

To validate the approach we perform the matching of Eq.~\eqref{eq:Chilinkin} to Eq.~\eqref{eq:DPD}~and~\eqref{eq:O.Bdecay}.
The equality of the first terms of both equations is clear.
For the second terms of both equations to be equal, it is required that
\begin{equation} \label{eq:magic.substitution}
D_{\lambda,\xi}^{1*}(\phi_+',\theta_+',0) e^{i\xi\alpha} =
\sum_{\lambda'} d_{\lambda\lambda'}^1(\wigner_{3(1)}^{1}) D_{\lambda',\xi}^{1*}(\phi_+, \theta_+,0),
\end{equation}
which would be valid if it holds for the rotation operators, \ie\xspace\footnote{
The angle $\wigner_{3(1)}^{1}$ implies clockwise rotation according to Eq.~\eqref{eq:flip.the.sign}.
To account for it, we take the angle to be negative, $\wigner_{3(1)}^{1} = -\wigner_{1(3)}^{1} < 0$.
}
\ie
\begin{equation}\label{eq:rotation.equality}
  R_z(\phi_+')R_y(\theta_+') R_z(\alpha) =
  R_y(\wigner_{3(1)}^{1}) R_z(\phi_+) R_y(\theta_+ ).
\end{equation}
The latter can be visualized by acting with the inverse rotations from Eq.~\eqref{eq:rotation.equality}
(in the order from left to right) on the system of particles $(\pi,K,B,\mu^+,\mu^-)$
in the $\psi$ rest frame obtained from the chain-$3$ shown in the left panel of Fig.~\ref{fig:alpha.angle}.
The application of the first Wigner rotation of the transformations on the right side of Eq.~\eqref{eq:rotation.equality}
is shown in the right panel of Fig.~\ref{fig:alpha.angle}.
The following two rotations bring the $\mu^+$ with from direction $(\theta_+,\phi_+)$ to the $z$ axis.
We note that $p_\pi$ stays in the $xz$ plane since it belongs to the blue muon plane.
The left-side transformations, applied to the left panel of Fig.~\ref{fig:alpha.angle}, already
align the direction of the $\mu^+$ with the $z$ axis directly with the first two rotations.
However, $\vec p_B$ is in the $xz$ plane (since it belongs to the blue plane) in that case.
The final azimuthal rotation $R_{z}(\alpha)$ on the left side of Eq.~\eqref{eq:rotation.equality}
brings the particle momenta to the same configuration as the right side does
since $\alpha$ is the difference of the azimuthal angles of $B$ and $\pi$ momenta
(note that $\vec p_{K^*} = \vec p_B$ in the $\psi$ rest frame; see also Fig.~14 in Ref.~\cite{Chilikin:2013tch})
in the configuration, where muons are aligned with the $z$ axis in the $\psi$ rest frame.

 % _|        _|    _|    _|_|_|  _|                                            _|                                                        _|
 % _|        _|    _|  _|        _|_|_|        _|_|_|      _|_|    _|_|_|    _|_|_|_|    _|_|_|    _|_|_|  _|    _|    _|_|_|  _|  _|_|  _|  _|
 % _|        _|_|_|_|  _|        _|    _|      _|    _|  _|_|_|_|  _|    _|    _|      _|    _|  _|    _|  _|    _|  _|    _|  _|_|      _|_|
 % _|        _|    _|  _|        _|    _|      _|    _|  _|        _|    _|    _|      _|    _|  _|    _|  _|    _|  _|    _|  _|        _|  _|
 % _|_|_|_|  _|    _|    _|_|_|  _|_|_|        _|_|_|      _|_|_|  _|    _|      _|_|    _|_|_|    _|_|_|    _|_|_|    _|_|_|  _|        _|    _|
 %                                             _|                                                      _|
 %                                             _|                                                      _|

\section{Relation to the LHCb pentaquark analysis} \label{sec:dec.Lb}
One of the most complicated amplitude analysis model has been applied to the decay \mbox{$\Lambda_b^0\to J/\psi[\to\mu^+\mu^-]\,p\,K^-$}~\cite{Aaij:2015tga}.
The amplitude was constructed using an isobar model with two chains, the $\Lambda$ states in the $pK$ subchannel (chain-$3$ in the discussion below), and $P_c$ chain (chain-$1$).
Each chain contains the $J/\psi\to\mu^+\mu^-$ decay, depending on the polar and azimuthal angles defined in the correspondent frames.
Using the notation of this paper, the LHCb model reads (\cf Eq.~(3,4,8) of Ref.~\cite{Aaij:2015tga}):
\begin{align} \label{eq:Pc.LHCb}
  \left(M^{\Lambda}_{\lambda,\xi}\right)_\text{LHCb} &=
  \sum_{s}^{\Lambda^*\to p K^-}\sum_{\tau,\mu}
  \sqrt{6}n_s\,
    D_{\Lambda,\tau-\mu}^{1/2*}(\phi_1,\theta_1,0)\,\HellH{0\to(23),1}_{\tau,\mu}\,\, \Ps{\sigma_1}\\ \nonumber
    &\hspace{3cm}
    \times D_{\tau,\lambda}^{s*}(\phi_{23},\theta_{23},0)\,\HellH{(23)\to 2,3}_{\lambda,0}\,\,
    D_{\mu\xi}^{1*}(\phi_+'', \theta_+,0)\,
    \HellH{1\to \mu^+\!,\mu^-}_{\lambda_+,\lambda_-}\\ \nonumber
  &\quad + \sum_{s}^{P_c\to J/\psi p}\sum_{\tau,\mu,\lambda'}
     \sqrt{6}n_s\,
     D_{\Lambda,\tau}^{1/2*}(\phi_3,\theta_3,0) \HellH{0\to(12),3}_{\tau,0}\,\,\Ps{\sigma_3}\\ \nonumber
     &\hspace{3cm}
     \times D_{\tau,\mu-\lambda}^{s*}(\phi_{12},\theta_{12},0)\,
     \HellH{(12)\to 1,2}_{\mu,\lambda'} \,
     d_{\lambda'\lambda}^{1/2}(\wigner_{3(1)}^{2})\,\,
     D_{\mu,\xi}^{1*}(\phi_+',\theta_+',0)\,
     \HellH{1\to \mu^+\!,\mu^-}_{\lambda_+,\lambda_-}\, e^{i\xi\alpha}.
\end{align}

To relate the $J/\psi$ decay angles in chain-$3$ to chain-$1$, Eq.~\eqref{eq:magic.substitution} is used.
In this case, the azimuthal angle between the $(J/\psi\,p\,K^-)$ and the $(J/\psi\,\mu^+\,\mu^-)$ planes is equal to $\phi_{23} + \phi_+''$
(see Fig.~16 in the Supplemental Material of Ref.~\cite{Aaij:2015tga}, where $\phi_{23} = \phi_K$, $\phi_+ = \phi_\mu$).
\begin{equation}
  D_{\mu\xi}^{1*}(\phi_+',\theta_+',0) e^{i\xi\alpha} =
  \sum_{\mu'} d_{\mu\mu'}^1(\wigner_{3(1)}^{1}) \, e^{i\mu' (\phi_{23} + \phi_+'')}\, d_{\mu'\xi}^{1}(\theta_+),
\end{equation}
where $\alpha$ is the difference of the azimuthal angles of $\Lambda_b$ and $p$ momenta
in the configuration when muons are aligned with $z$ axis in the $\psi$ rest frame, analogous to the $B$ decay in Appendix~\ref{sec:dec.B}.

To separate the overall rotation we transform the Wigner $D$-functions for both chain-$1$ and chain-$3$:
For the chain-$1$, factoring $D^{J*}_{\Lambda,\nu}(\phi_1,\theta_1,\phi_{23})$ is simply:
\begin{equation} \label{eq:DD}
  \sum_{\tau} D^{J*}_{\Lambda,\tau-\lambda_1}(\Omega_1) D^{s*}_{\tau,\lambda_2-\lambda_3}(\Omega_{23}) =
  \sum_\nu D^{J*}_{\Lambda,\nu}(\phi_1,\theta_1,\phi_{23}) \left[
    \sum_\tau \delta_{\nu,\tau-\lambda_1} e^{i\lambda_1 \phi_{23}} d^{s}_{\nu,\lambda_2-\lambda_3}(\theta_{23})
  \right]
\end{equation}
For the chain-$3$, the decomposition requires an additional step as follows:
\begin{align}
  \sum_\tau D_{\Lambda,\tau}^{1/2*}(\phi_3,\theta_3,0) D_{\tau,\mu-\lambda}^{s*}(\phi_{12},\theta_{12},0) &=
  \sum_\tau D_{\Lambda,\tau}^{1/2*}(\phi_3,\theta_3,\phi_{12}) d_{\tau,\mu-\lambda}^{s}(\theta_{12})\\\nonumber
  &=\sum_{\nu,\tau} D_{\Lambda,\nu}^{1/2*}(\phi_1,\theta_1,\phi_{23}) d_{\nu,\tau}^{1/2}(\hat{\theta}_{3(1)}) d_{\tau,\mu-\lambda}^{s}(\theta_{12}),
\end{align}
where we used $R(\phi_3,\theta_3,\phi_{12}) = R(\phi_1,\theta_1,\phi_{23}) R_y(\hat{\theta}_{3(1)})$.

With all substitutions, the expression in Eq.~\eqref{eq:Pc.LHCb} is transformed into the desired form:
\begin{align}
  \left(M^{\Lambda}_{\lambda,\xi}\right)_\text{LHCb} &=
  \sum_{\nu,\mu} D_{\Lambda,\nu}^{J*}(\phi_1,\theta_1,\phi_{23})\\ \nonumber
  &\quad\times\bigg(
\sum_{s}^{\Lambda^*\to p K}\sum_\tau
  \sqrt{2}n_s\,
  d_{\nu,\tau-\mu}^{1/2}(0)\,\HellH{0\to(23),1}_{\tau,\mu}
  \,\,\Ps{\sigma_1}\,\,
  d_{\tau,\lambda}^{s}(\theta_{23})\,\HellH{(23)\to 2,3}_{\lambda,0}\\  \nonumber
&\quad\quad + \sum_{s}^{P_c\to J/\psi p}\sum_{\tau,\mu',\lambda'}
    \sqrt{2}n_s\,
    d_{\nu,\tau}^{1/2}(\hat\theta_{3(1)})\,\HellH{0\to(12),3}_{\tau,0}
    \,\,\Ps{\sigma_3}\,\,
    d_{\tau,\mu'-\lambda'}^{s}(\theta_{12})\,\HellH{(12)\to 1,2}_{\mu',\lambda'} \,\,
    d_{\lambda'\lambda}^{1/2}(\wigner_{3(1)}^{2})\,\,
    d_{\mu'\mu}^1(\wigner_{3(1)}^{1})
  \bigg)\,\,\\  \nonumber
  &\quad
  \times
  \sqrt{3}\,
  e^{i\mu(\phi_{23}+\phi_+'')}\,d_{\mu\xi}^{1}(\theta_+)\,
  \HellH{1\to \mu^+\!,\mu^-}_{\lambda_+,\lambda_-},
\end{align}
where the form of the amplitude matches Eq.~\ref{eq:Lb.rot} with $\phi_+ = \phi_{23}+\phi_+''$.

The last step is to examine is the helicity state of the proton.
It is defined in the $\Lambda^*$ rest frame in Eq.~\ref{eq:Pc.LHCb}, while the particle-$0$ rest frame is used in the conventions of Eq.~\eqref{eq:O.Ldecay}.
Hence,
\begin{equation} \label{eq:M.LHCb.vs.M}
  \left( M^{\Lambda}_{\lambda;\lambda_+\lambda_-} \right)_{\text{Eq.}~(\ref{eq:Lb.rot}-\ref{eq:O.Ldecay})} =
    \sum_{\lambda'} \left(M^{\Lambda}_{\lambda',\lambda_+-\lambda_-}\right)_\text{LHCb} d_{\lambda'\lambda}^{1/2}(\wigner^{2}_{1(0)}).
\end{equation}
Using the sum rule from Eq.~\eqref{eq:wigner.123},
$$
\sum_{\lambda''} d_{\lambda'\lambda''}^{1/2}(\wigner_{3(1)}^{2}) d_{\lambda''\lambda}^{1/2}(\wigner^{2}_{1(0)}) = d_{\lambda'\lambda}^{1/2}(\wigner_{3(0)}^{2}) = d_{\lambda'\lambda}^{1/2}(\wigner_{3(2)}^{2}),
$$
we complete the proof of the equivalence of the LHCb formalism (Eq.~\eqref{eq:Pc.LHCb}) and the one we presented in this paper (Eqs.~(\ref{eq:Lb.rot}-\ref{eq:O.Ldecay})).

Curiously, the Wigner $d$-function in Eq.~\eqref{eq:M.LHCb.vs.M} does not change the differential distributions when the squared matrix element is summed over the proton helicity,
it is canceled in the summation, due to the relation:
$$
\sum_{\lambda} d_{\lambda'\lambda}^{1/2}(\wigner^{2}_{1(0)}) d_{\lambda''\lambda}^{1/2}(\wigner^{2}_{1(0)}) = \delta_{\lambda'\lambda''}.
$$

% \begin{align} \label{eq:M.LHCb.vs.M}
%   \sum_{\lambda}\left|\left( M^{\Lambda}_{\lambda;\lambda_+\lambda_-} \right)_{\text{Eq.}~(\ref{eq:Lb.rot}-\ref{eq:O.Ldecay})}\right|^2
%   &=
%   \sum_{\lambda}\sum_{\lambda'\lambda''} \left(M^{\Lambda}_{\lambda',\lambda_+-\lambda_-}\right)_\text{LHCb} d_{\lambda'\lambda}^{1/2}(\wigner^{2}_{1(0)})
%   \left(M^{\Lambda}_{\lambda'',\lambda_+-\lambda_-}\right)_\text{LHCb} d_{\lambda''\lambda}^{1/2}(\wigner^{2}_{1(0)})\\\nonumber
%   &=
%   \sum_{\lambda'}\left|\left(M^{\Lambda}_{\lambda',\lambda_+-\lambda_-}\right)_\text{LHCb} \right|^2
% \end{align}

\section{Azimuthal Wigner rotations} \label{sec:complex.wigner.D}
The conventional helicity formalism for a three-body decay amplitude
in the isobar model requires a sum of three truncated partial-wave series over four angles in the space-fixed CM.
The amplitudes carrying helicity indices, however, need to be added with care to make sure that the spin-quantization axes of all particles
are the same in the three terms. The framework proposed in the main text provides a simple approach to ensure it.
Nevertheless, the consistency can also be achieved by combined in the space-fixed CM rather then in the aligned CM. To match the quantization axes of the different decay chains, one needs to add an extra azimuthal rotation to the polar $R_y(\wigner_{j(0)}^i)$ discussed in the main text. The correspondent matrix element thus has an extra complex phase.
%
% Nevertheless, the consistency can be also achieved by using the complex Wigner rotations for the terms combined in the space-fixed CM rather then in the aligned CM.
% In this case, in addition to the rotation $R_y(\wigner_{j(0)}^i)$ discussed in the main text, the Wigner rotations have complex phase which is understood by
%a need of an azimuthal rotation to match quantization axes of different decay chains.
Here we derive this extra rotation algebraically by imposing the factorization from Eq.~\eqref{eq:proposal}.

The decay $\Lambda_c\to p K\pi$ is used as an example, since all three decay chains are included in the construction, thus making the case general.
Three decay amplitudes are built using Eq.~\eqref{eq:h1} and Eq.~\eqref{eq:h2}.
% \begin{subequations}
\begin{align}
  M_{\Lambda\lambda}^{(1)} &= \sqrt{2}\sum_{s}^{K^*\to K \pi}\sum_\tau
            n_s\,
            D^{1/2*}_{\Lambda,\tau-\lambda}(\phi_1,\theta_1,0) \HellH{0\to(23),1}_{\tau,\lambda} \,
            \Props{\sigma_1}{}\,
            D^{s*}_{\tau,0}(\phi_{23}, \theta_{23}, 0) \HellH{(23)\to 2,3}_{0,0},\\ \nonumber
  M_{\Lambda\lambda}^{(2)} &= \sqrt{2}\sum_{s}^{\Delta\to \pi p}\sum_{\tau}
            n_s\,
            D^{1/2*}_{\Lambda,\tau}(\phi_2,\theta_{2},0) \HellH{0\to(31),2}_{\tau,0}\,
            \Props{\sigma_2}{}\,
            D^{s*}_{\tau,-\lambda}(\phi_{31},\theta_{31},0) \HellH{(31)\to 3,1}_{0,\lambda},\\ \nonumber
  M_{\Lambda\lambda}^{(3)} &= \sqrt{2}\sum_{s}^{\Lambda\to p K}\sum_{\tau}
            n_s\,
            D^{1/2*}_{\Lambda,\tau}(\phi_3, \theta_{3},0)\, \HellH{0\to(12),3}_{\tau,0}\,
            \Props{\sigma_3}{}\,
            D^{s*}_{\tau,\lambda}(\phi_{12},\theta_{12},0) \HellH{(12)\to 1,2}_{\lambda,0}.
\end{align}
% \end{subequations}
More details on the construction can be found in Sec.~\ref{sec:Lc}.

The factorization can be ensured once the overall rotation is factored out from each term $M^{(k)}$, $k=1,2,3$.
We use the relation:
% \begin{subequations} \label{eq:transformations}
\begin{align} \nonumber
    D^{1/2*}_{\Lambda,\tau-\lambda}(\phi_1,\theta_1,0) D^{s*}_{\tau,0}(\phi_{23}, \theta_{23}, 0) &=
    D^{1/2*}_{\Lambda,\tau-\lambda}(\phi_1,\theta_1,\phi_{23}) d^{s}_{\tau,0}(\theta_{23}) e^{i\lambda\phi_{23}} =
    \sum_\nu D^{1/2*}_{\Lambda,\nu}(\phi_1,\theta_{1},\phi_{23}) \delta_{\nu,\tau-\lambda} d^{s}_{\tau,0}(\theta_{23}) e^{i\lambda\phi_{23}},  \\ \nonumber
    D^{1/2*}_{\Lambda,\tau}(\phi_2,\theta_{2},0) D^{s*}_{\tau,-\lambda}(\phi_{31},\theta_{31},0) & =
    D^{1/2*}_{\Lambda,\tau}(\phi_2,\theta_{2},\phi_{31}) d^{s}_{\tau,-\lambda}(\theta_{31}) =
    \sum_\nu D^{1/2*}_{\Lambda,\nu}(\phi_1,\theta_{1},\phi_{23}) d^{1/2}_{\nu,\tau}(\hat\theta_{2(1)}) d^{s}_{\tau,-\lambda}(\theta_{31}), \\  \label{eq:transformations}
    D^{1/2*}_{\Lambda,\tau}(\phi_3, \theta_{3},0) D^{s*}_{\tau,\lambda}(\phi_{12},\theta_{12},0) &=
    D^{1/2*}_{\Lambda,\tau}(\phi_3, \theta_{3},\phi_{12}) d^{s}_{\tau,\lambda}(\theta_{12}) =
    \sum_\nu D^{1/2*}_{\Lambda,\nu}(\phi_1,\theta_{1},\phi_{23}) d^{1/2}_{\nu,\tau}(\hat\theta_{3(1)})  d^{s}_{\tau,\lambda}(\theta_{12}).
\end{align}
% \end{subequations}
In the left equalities, we move the phase due to azimuthal rotation $R_z(\phi_{ij})$
from the second to the first Wigner matrix.
The right equalities rely on the identity discussed in Sec.~\ref{sec:dalitz.plot.decomposition}:
$$
R_{z}(\phi_k) R_y(\theta_k) R_{z}(\phi_{ij}) = R_{z}(\phi_1) R_y(\theta_1) R_{z}(\phi_{23}) R_y(\hat{\theta}_{k(1)}).
$$
The phase factor in the first line of Eq.~\eqref{eq:transformations} indicates that
the particle-$1$ quantization axis
in the chain-$1$ and in the chain-$2$ and $3$ differ by an azimuthal angle. Hence, this must be included in the Wigner rotation
that matches the quantization axes.
\begin{align}
    M_{\Lambda\lambda} &= M_{\Lambda\lambda}^{(1)} +
        \sum_{\lambda'} M_{\Lambda\lambda'}^{(2)} \, D_{\lambda'\lambda}^{1/2*}(0,\wigner_{2(1)}^{1},\phi_{23}) +
        \sum_{\lambda'} M_{\Lambda\lambda'}^{(3)} \, D_{\lambda'\lambda}^{1/2*}(0,\wigner_{3(1)}^{1},\phi_{23})\\ \nonumber
        &= D^{1/2*}_{\Lambda,\nu}(\phi_1,\theta_{1},\phi_{23}) O_{\lambda}^{\nu}(\{\sigma_i\}) e^{i\lambda\phi_{23}}
\end{align}
with $O_{\lambda}^{\nu}$ given by Eq.~\eqref{eq:O.Lc}. The overall phase is unobservable.

\bibliographystyle{apsrev4-1}
\bibliography{dalitzplot}

\end{document}